# Physics-Infused Reduced-Order Modeling for Analysis of Multi-Layered Hypersonic Thermal Protection Systems

Carlos A. Vargas Venegas * and Daning Huang†
*The Pennsylvania State University, University Park, Pennsylvania, 16801*

Patrick Blonigan and John Tencer‡
*Sandia National Laboratories, Albuquerque, New Mexico, 87123*

This work presents a physics-infused reduced-order modeling (PIROM) framework for efficient and accurate prediction of transient thermal behavior in multi-layered hypersonic thermal protection systems (TPS). The PIROM architecture integrates a reduced-physics backbone, based on the lumped-capacitance model (LCM), with data-driven correction dynamics formulated via a coarse-graining approach rooted in the Mori-Zwanzig formalism. While the LCM captures the dominant heat transfer mechanisms, the correction terms compensate for residual dynamics arising from higher-order non-linear interactions and heterogeneities across material layers. The proposed PIROM is benchmarked against two non-intrusive reduced-order models (ROMs): Operator Inference (OpInf) and Neural Ordinary Differential Equations (NODE). The PIROM consistently achieves errors below $1\%$ for a wide range of extrapolative settings involving time- and space-dependent boundary conditions and temperature-varying material property perturbations. In contrast, OpInf exhibits moderate degradation, and NODE suffers substantial loss in accuracy due to its lack of embedded physics. Despite higher training costs, PIROM delivers online evaluations of two orders of magnitude faster than the full-order model. These results demonstrate that PIROM effectively reconciles the trade-offs between accuracy, generalizability, and efficiency, providing a robust framework for thermal modeling of TPS under diverse operating conditions.

## Nomenclature

**Roman Symbols**

$\mathbf{A}, \mathbf{A}_i, \mathbf{B}, \mathbf{B}_{ij}$    System matrices and their block-wise components

$c_p$      Specific heat

---

*Graduate Student Research Assistant, Aerospace Engineering, University Park, and AIAA Student Member
†Assistant Professor, Aerospace Engineering, University Park, and AIAA Member
‡Insert Job Title, Department Name, Address/Mail Stop, and AIAA Member Grade (if any) for third author.

| Symbol | Description |
|---|---|
| $\mathcal{D}$ | Dataset |
| $d$ | Dimension of the domain |
| $E_i$ | The $i$th element |
| $\mathcal{E}$ | Collection of edges |
| $e_{ij}$ | Boundary shared by elements $i$ and $j$ |
| $\mathbf{f}, \mathbf{f}_i$ | Forcing vector and the $i$th component |
| $\mathcal{J}$ | Objective function |
| $\mathbf{k}, k$ | Thermal conductivity |
| $\ell$ | Cost function |
| $M$ | Number of elements |
| $\mathcal{N}_i$ | Collection of neighbors for element $i$ |
| $N$ | Number of layers in TPS |
| $N_p$ | Number of steps in a trajectory |
| $N_s$ | Number of high-fidelity trajectories |
| $\mathcal{P}, \mathcal{Q}$ | Projection and Residual operators |
| $P$ | Number of basis functions of an element |
| $q$ | Heat flux |
| $R$ | Thermal resistance |
| $\mathbf{r}^{(1)}, \mathbf{r}^{(2)}$ | Resolved and Unresolved dynamics |
| $T$ | Temperature |
| $t, t_0, t_f$ | Time, initial time, final time |
| $\mathbf{u}, \mathbf{u}_i$ | Full state vector and the states of element $i$ |
| $x$ | Spatial coordinate |
| $\mathbf{y}$ | State vector of PIROM |
| **z** | Observables |

**Greek Symbols**

| | |
|---|---|
| $\boldsymbol{\beta}$ | Hidden states |
| $\Gamma$ | Domain boundary of partial differential equation |
| $\epsilon$ | Tolerance in training |
| $\boldsymbol{\Theta}$ | Learnable parameters in machine learning model |
| $\kappa, \tilde{\kappa}$ | Memory kernels |
| $\rho$ | Density |
| $\sigma$ | Penalty factor in Discontinuous Galerkin |
| $\boldsymbol{\Phi}, \boldsymbol{\Phi}^+$ | Projection matrix and its pseudo-inverse |
| $\boldsymbol{\phi}_i, \phi_p^i$ | Basis function of element $i$, and its $p$th component |
| $\boldsymbol{\varphi}_i^j$ | Weight vector mapping states in element $i$ to average temperature $j$ |
| $\boldsymbol{\xi}$ | Parameters for operating conditions |
| $\Omega$ | Domain of partial differential equation |

**Other symbols**

| | |
|---|---|
| $\bar{\Box}$ | Coarse-grained quantity |
| $\Box_{ij}$ | Quantity related to elements $i$ and $j$ |
| $\Box_{HF}$ | Quantities computed using the high-fidelity solver |
| $\Box_q$ | Quantities related to Neumann boundary condition |
| $\Box_T$ | quantities related to Dirichlet boundary condition |

**Acronyms**

| | |
|---|---|
| BC | Boundary Condition |
| DG | Discontinuous Galerkin |
| FEM | Finite-Element Method |

FOM    Full-Order Model

IPG    Interior-Penalty Galerkin

LCM    Lumped-Capacitance Model

MZ    Mori-Zwanzig

NN    Neural Network

NODE    Neural Ordinary Differential Equation

NRMSE    Normalized Root-Mean Squared Error

OOD    Out of Distribution

OpInf    Operator Inference

PIROM    Physics-Infused Reduced-Order Modeling

ROM    Reduced-Order Model

RPM    Reduced-Physics Model

TPS    Thermal Protection System

## I. Introduction

Thermal protection systems (TPS) play a critical role in ensuring safety and optimal performance of hypersonic vehicles operating in extreme environments such as atmospheric re-entry or long-range hypersonic cruise. The purpose of the TPS is to shield interior structures from excessive heating rates that may exceed thousands of Watts per square centimeter, depending on the speed, altitude, and trajectory of the space vehicle. For example, the Space Shuttle's TPS – consisting primarily of reusable silica-based ceramic tiles on the fuselage and reinforced carbon-carbon (RCC) panels on the leading edges – was crucial in protecting the orbiter's aluminum airframe from the surface temperatures reaching approximately 1,650 °C during re-entry into the Earth's atmosphere [1]. Advanced hypersonic aircraft, such as the X-51 Waverider and X-43 research plane, employed ablative ceramic matrix composites and heat-resistant alloys designed to withstand temperature above 1,200 °C at speeds approaching Mach 7 [2, 3]. Other applications, such as the Mars Science Laboratory (MSL) rover, used a heat shield composed of phenolic impregnated carbon ablator (PICA) to protect the rover from the intense heat generated during the entry, descent, and landing (EDL) phase on Mars [4]. The examples above demonstrate the critical role of TPS in ensuring the survivability and optimal performance of hypersonic vehicles, underscoring the need for meticulous analysis, design, and rapid manufacturing processes that adhere to strict time and budget constraints.

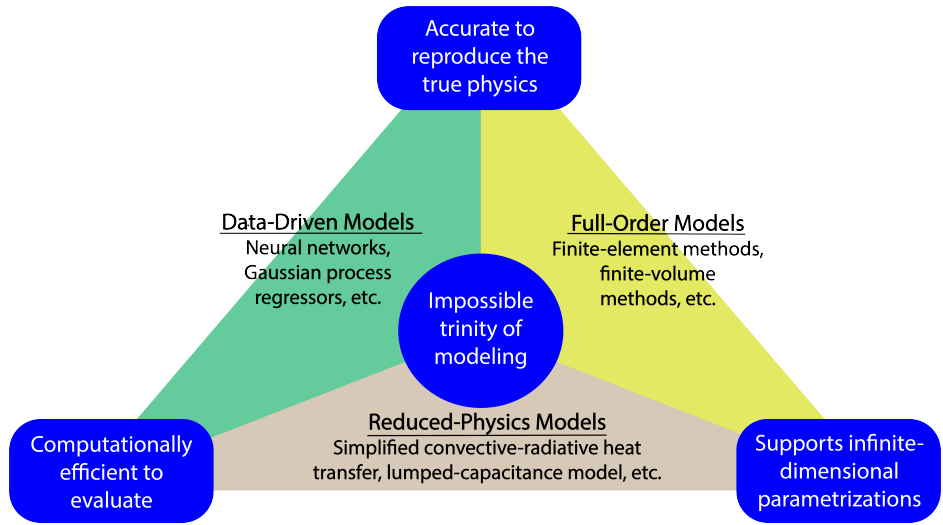

Fig. 1   The impossible trinity of modeling in the context of TPS.

Nevertheless, designing TPS for complex vehicle geometries – e.g., those with wing-body blending, control surfaces, or highly curved forebodies – introduces geometry-induced complexities in both fluid flow and structural layouts. The presence of such geometric features can lead to highly non-uniform heating rates, shock-layer interactions, and boundary-layer transition, which can significantly affect the performance and survivability of the TPS. Furthermore, accounting for both in-plane and through-thickness variability of material properties, while resolving the difference in local geometries, adds to the computational burden. The high-dimensionality of the aero-thermo-structural domains, together with the extensive exploration required in the design space, leads to computationally prohibitive demands when relying on high-fidelity simulations – such as large-scale computational fluid dynamics (CFD) and finite-element methods (FEMs) for fluid and structural analysis, respectively – due to their immense cost in memory and run time.

Reduced-order models (ROMs) have emerged as a promising approach to address the computational challenges associated with thermodynamic modeling of TPS. Ideally, a ROM should be (1) accurate to reproduce high-fidelity solutions, (2) support continuous or infinite-dimensional design parameters such as geometrical shapes and material distributions, and (3) be computationally efficient to evaluate to allow for fast turnaround time in design optimization. However, the above three capabilities usually form an impossible trinity of modeling, as illustrated in Fig. 1; building a ROM that achieves any two capabilities sacrifices the third.

The impossible trinity poses a significant challenge in the development of ROMs for the multi-disciplinary analysis and optimization of TPS. More specifically, full-order models (FOMs), such as finite-element methods (FEMs), offer high accuracy and robust handling of diverse design space, but they are computationally expensive to run. Reduced-physics models (RPMs) – such as simplified convective-radiative heat transfer or engineering correlations – achieve efficiency and broad applicability by ignoring certain non-linear or small-scale effects, however, RPMs sacrifice accuracy for complex thermal boundary conditions or material behaviors. Lastly, data-driven ROMs, such as Gaussian Process Regressors [5] and neural ordinary differential equations (NODE) [6], do provide accurate and computationally efficient approximations of highly non-linear input-output for the temperature distributions over the TPS layers; nonetheless, these data-centric approaches often demand extensive high-fidelity data for training, do not necessarily satisfy fundamental physical constraints or conservation laws, and thus do not generalize well to design space outside the training data.

The landscape of ROM techniques is rapidly changing over time, and some developments may have brought ROMs closer to solving the impossible trinity of modeling. These ROMs are typically divided into intrusive and non-intrusive approaches. In traditional projection-based methods, such as proper-orthogonal decomposition with Galerkin projections (POD-G) [7–10], a reduced-form representation of the full-order solution is introduced into the FOM to reduce problem dimensionality and preserve the governing physics [11]. Although successful in thermal modeling of ablative TPS systems [12, 13], intrusive methods are only applicable when FOM operators are available, which is not practical in many engineering scenarios where high-fidelity solvers are treated as black boxes.

In contrast, non-intrusive ROMs circumvent the need for direct access to the FOM equations by learning reduced dynamics from high-fidelity simulation or measurement data. Towards the solution of the impossible trinity of modeling, recent advances in non-intrusive reduced-order modeling have emphasized physics-informed approaches; such approaches incorporate physical knowledge in the ROM so as to increase generalizing capability to unknown operating scenarios, particularly when training data are scarce. Physics-informed neural networks (PINNs) [14–16] incorporate governing laws (e.g., conservation of mass, momentum, energy) into the loss function during training, improving generalizability and reducing the reliance on large datasets [17]. For example, when predicting the stagnation point heat flux over a blunt-nosed body [18], the PINN achieved a predictive accuracy of 4.5% for parameters involving geometries and operating conditions outside the training datasets – this is in contrast to the purely data-driven model, whose error nearly doubled. Physics-Informed DeepONets (PIDeepONets) [19–21] extended this idea to operator learning, capturing mappings between function spaces under physics constraints. For example, in Ref. [22], it is shown that by including the continuity equation in the objective function, the PIDeepONet can extrapolate chemically-reacting hypersonic wave predictions with an accuracy of 5% for wave perturbations and space-time domains outside the training datasets; the predictions are done for cases where the full-field data is partially known.

Another instance of non-intrusive modeling is Operator inference (OpInf) [23]. The physics-informed nature of the OpInf provides a balance between data-driven flexibility and physics-based structure, and is termed as a "glass-box" approach, as the targeted dynamics are known via the governing PDEs that define the problem of interest, but the inner workings of the high-fidelity solver are unknown. The OpInf projects the governing PDE dynamics onto a reduced basis constructed from high-fidelity snapshot data; the result is a ROM based on the structure of the projected PDE, typically in the form of polynomials. The learning of the operators is performed in the projected space, using a least-squares approach on high-fidelity state data [11]. The OpInf has demonstrated success in rocket combustion [23], TPS thermal modeling [24], and fluid dynamics [23], achieving over three orders of magnitude of computational speed-ups relative to the CFD high-fidelity solvers.

Nonetheless, these methods also exhibit limitations: PINNs and PI-DeepONets only weakly enforce physical laws during training and potentially require large datasets to achieve high accuracy. Meanwhile, OpInf models face challenges when parametric dependency is needed. Due to the model form construction, the parameters need to appear as explicit non-linear features in the OpInf model. The numbers of such features would grow exponentially as more parameters and more non-linear relations are involved, e.g., cubic or quartic operators, which poses a challenge in learning such model from data; this renders OpInf non-ideal for vast design space explorations under scarce high-fidelity data.

This work presents a novel physics-infused reduced-order modeling (PIROM) approach [25–28], a non-intrusive hybrid framework that combines the strengths of physics-based models with machine learning, to formulate and train a ROM for parametrized non-linear dynamical systems. The PIROM methodology is demonstrated for the transient thermodynamics analysis of multi-layered TPS in a high-speed vehicle. The physics component is the lumped capacitance model (LCM), a classical theoretical model of heat conduction and convection, that captures physically-sound, though not accurate, transient temperature dynamics in the multi-layered TPS. The LCM captures arbitrary design parameters in boundary conditions and material properties and can account for a level of non-linearity such as temperature-dependent material properties. The LCM is extended with a data-driven hidden dynamics to account for the missing physics in the LCM, which are learned from high-fidelity data. The hidden dynamics enables higher predictive accuracy of the PIROM when subjected to complex boundary conditions and new material properties. Through this hybrid approach, the PIROM aims to solve the impossible trinity of modeling by leveraging the generalizability and computational efficiency of RPMs, while incorporating the accuracy and adaptability of data-driven extensions.

Moreover, in prior work on PIROM, the hidden dynamics for correction is devised in a heuristic manner, e.g., by introducing corrections to semi-empirical coefficients in a boundary layer model [27]; such heuristic approach might not be applicable to other applications, such as the TPS problem. This study adopts earlier work on coarse-graining modeling of non-linear dynamics [29] to rigorously derive the model form for a PIROM using the Mori-Zwanzig formalism [30–33]. For the TPS problem, this formulation produces a sufficiently simple model form while maintaining the physical consistency of PIROM. More importantly, the formulation provides a general methodology for developing PIROMs for other problems.

In sum, the objectives of this work are as follows:

1) Formulate the PIROM for the transient thermodynamics of the multi-layered TPS through a systematic coarse-graining procedure based on the Mori-Zwanzig formalism.
2) Perform the convergence study of the PIROM, especially its convergence to the full-order solution as the model size increases.

3) Benchmark the accuracy, generalizability, and efficiency of the PIROM with respect to the other representative non-intrusive ROMs, including OpInf and NODE.

## II. Modeling of Thermal Protection Systems

This section formulates the heat conduction problem for a generic TPS geometry, and provides two different but mathematically-connected solution strategies: (1) a high-fidelity full-order model (FOM) based on a discontinuous Galerkin finite element method, and (2) a low-fidelity reduced-physics model (RPM) based on a lumped capacitance model (LCM). The FOM is computationally expensive but provides the highest fidelity, while the RPM is computationally efficient but has low predictive fidelity; both models are amenable to high-dimensional design variables. The RPM is used in the subsequent sections for deriving the PIROM.

### A. Governing Equations

Consider a generic domain $\Omega \subset \mathbb{R}^d$, $d = 2$ or $3$, with prescribed heat flux $q_b$ on boundary $\Gamma_q$ (i.e., Neumann boundary condition) and prescribed temperature $T_b$ on boundary $\Gamma_T$ (i.e., Dirichlet boundary condition), where $\partial \Omega = \Gamma_q \cup \Gamma_T$. The transient heat conduction dynamics is governed by the following equation,

$$\rho c_p \frac{\partial T}{\partial t} - \nabla \cdot (\mathbf{k} \nabla T) = 0, \ x \in \Omega \tag{1a}$$

$$-\mathbf{k} \nabla T \cdot \mathbf{n} = q_b(x, t), \ x \in \Gamma_q \tag{1b}$$

$$T(x, t) = T_b(x, t), \ x \in \Gamma_T \tag{1c}$$

$$T(x, 0) = T_0(x), \ x \in \Omega, \tag{1d}$$

where the density $\rho$ is constant, while heat capacity $c_p$, and thermal conductivity $\mathbf{k} \in \mathbb{R}^{d \times d}$, are temperature-dependent. The TPS domain is divided into $N$ non-overlapping components $\{\Omega_i\}_{i=1}^N$, as illustrated in Fig 2 for $N = 2$. The $i$-th component $\Omega_i$ is associated with a different set of material properties $(\rho_i, c_{p,i}, \mathbf{k}_i)$, that are assumed to be continuous within one component, and can be discontinuous across two neighboring components.

### B. Full-Order Model: Finite Element Method

To obtain the full-order numerical solution, the governing equation is spatially discretized using variational principles of Discontinuous Galerkin (DG) to result in a high-dimensional system of ordinary differential equations (ODEs). Furthermore, the DG model is written in an element-wise form, which is beneficial for subsequent derivations for the lower-order models. Note that the choice of DG approach here is mainly for theoretical convenience in the subsequent coarse-graining formulation. In Sec. IV, the full-order TPS solution is computed using standard FEM instead, and the equivalence between DG and standard FEM is noted upon their convergence.

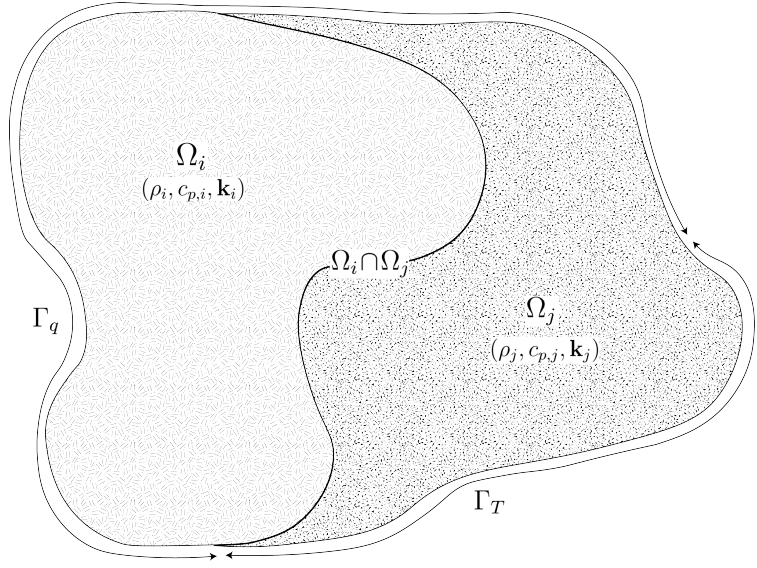

**Fig. 2  Composite structure inside TPS.**

*1. Domain Discretization*

Consider a conforming mesh partition of the domain, as shown in Fig. 2, where each element belongs to one and only one component. Denote the collection of all $M$ elements as $\{E_i\}_{i=1}^{M}$. To ease the description of the DG model, a graph structure is employed. The elements are treated as vertices, the set of which is denoted $\mathcal{V} = \{m\}_{m=1}^{M}$. Two neighboring elements, $E_i$ and $E_j$, are connected by an edge $(i, j)$, and the shared boundary between $E_i$ and $E_j$ is denoted $e_{ij}$. The collection of all edges are denoted $\mathcal{E}$, and $\mathcal{G} = (\mathcal{V}, \mathcal{E})$ is referred to as a graph. In the graph, the edges are undirected, meaning if $(i, j) \in \mathcal{E}$ then $(j, i) \in \mathcal{E}$. Furthermore, denote the neighbors of the $i$th element as $\mathcal{N}_i = \{j | (i, j) \in \mathcal{E}\}$. Lastly, for the ease of notation, introduce two special indices: $T$ for the boundary of an element that overlaps with the Dirichlet boundary condition, and similarly $q$ for the Neumann boundary condition.

*2. Weak Form of Discontinuous Galerkin Method*

Choosing appropriate basis functions $v$ and $w$ and using the Interior Penalty Galerkin (IPG) scheme [34], the variational bilinear form for Eq. (1) is,

$$\sum_{i=1}^{M} a_{\epsilon,i}(v, w) = \sum_{i=1}^{M} L_i(v), \tag{2}$$

where $\epsilon$ is a user-specified parameter, and

$$a_{\epsilon,i}(v, w) = \int_{E_i} \left(\rho c_p v \frac{\partial w}{\partial t} + \mathbf{k}\nabla v \cdot \nabla w\right) d\Omega$$
$$- \sum_{j \in \mathcal{N}_i \cup \{T\}} \int_{e_{ij}} \{\mathbf{k}\nabla v \cdot n\}[w] d\Gamma + \epsilon \sum_{j \in \mathcal{N}_i \cup \{T\}} \int_{e_{ij}} \{\mathbf{k}\nabla w \cdot n\}[v] d\Gamma + \sigma \sum_{j \in \mathcal{N}_i \cup \{T\}} \int_{e_{ij}} [v][w] d\Gamma \quad (3a)$$

$$L_i(v) = \epsilon \sum_{j \in \mathcal{N}_i \cup \{T\}} \int_{e_{ij}} (\mathbf{k}\nabla w \cdot n) T_b d\Gamma + \int_{e_{iq}} v q_b d\Gamma + \sigma \int_{e_{iT}} v T_b d\Gamma. \quad (3b)$$

In the bi-linear form above, the notations of [] and {} are respectively the jumps and averages at a boundary $e$ shared by two elements, $E_1$ and $E_2$,

$$[u] = u|_{E_1} - u|_{E_2}, \quad \{u\} = \frac{1}{2}(u|_{E_1} + u|_{E_2}), \quad \text{for } x \in e = E_1 \cap E_2. \quad (4)$$

Furthermore, in the bi-linear form, the terms associated with $\sigma$ are introduced to enforce the Dirichlet boundary condition; $\sigma$ is a penalty factor whose value can depend on the size of an element. Depending of the choice of $\epsilon$, the bi-linear form corresponds to symmetric IPG ($\epsilon = -1$), non-symmetric IPG ($\epsilon = 1$), and incomplete IPG ($\epsilon = 0$). All these schemes are consistent with the original PDE and have similar convergence rate with respect to mesh size. In the following derivations, the case $\epsilon = 0$ is chosen for the sake of simplicity.

*3. Discontinuous Galerkin Model*

Next, the DG-based full-order model is written in an element-wise form. For the $i$-th element, use a set of $P$ trial functions, such as polynomials, to represent the temperature distribution,

$$T(x, t) = \sum_{p=1}^{P} \phi_p^i(x) u_p^i(t) \equiv \boldsymbol{\phi}_i(x)^\top \mathbf{u}_i(t). \quad (5)$$

Without loss of generality, the trial functions are assumed to be orthogonal, so that $\int_{E_i} \phi_p^i \phi_q^i d\Omega = |E_i| \delta_{pq}$, where $|E_i|$ is the area ($d = 2$) or volume ($d = 3$) of the element and $\delta_{pq}$ is Kronecker delta. Furthermore, for simplicity, choose $\phi_1^i = 1$; by orthogonality, $\int_{E_i} \phi_p^i d\Omega = 0$ for $p > 1$. Under this choice of basis, $u_1^i$ is simply the average temperature of the element $E_i$, denoted $\bar{u}_i$.

Using test functions same as trial functions, the dynamics of $\mathbf{u}_i$ is obtained by evaluating the element-wise bilinear forms

$$a_{\epsilon,i}(\phi_p^i, w) = L_i(\phi_p^i), \quad p = 1, 2, \cdots, P. \quad (6)$$

The above procedure produces,

$$\mathbf{A}_i \dot{\mathbf{u}}_i = \mathbf{B}_i \mathbf{u}_i + \sum_{j \in \mathcal{N}_i \cup \{T\}} \left( \mathbf{B}_{ij}^i \mathbf{u}_i + \mathbf{B}_{ij}^j \mathbf{u}_j \right) + \mathbf{f}_i, \qquad (7)$$

where, for $k, l = 1, 2, \cdots, P$,

$$[\mathbf{A}_i]_{kl} = \int_{E_i} \rho c_p \phi_k^i \phi_l^i d\Omega, \quad [\mathbf{B}_i]_{kl} = \int_{E_i} (\nabla \phi_k^i) \cdot (\mathbf{k} \nabla \phi_l^i) d\Omega, \quad [\mathbf{f}_i]_k = \int_{e_{iq}} \phi_k^i q_b d\Gamma + \sigma \int_{e_{iT}} \phi_k^i T_b d\Gamma, \qquad (8a)$$

$$[\mathbf{B}_{ij}^i]_{kl} = \begin{cases} -\int_{e_{ij}} \{\mathbf{k} \nabla \phi_k^i \cdot n\} \phi_l^i + \sigma [\phi_k^i] \phi_l^i d\Gamma & j \in \mathcal{N}_i \cup \{T\} \\ 0 & \text{otherwise} \end{cases}, \qquad (8b)$$

$$[\mathbf{B}_{ij}^j]_{kl} = \begin{cases} \int_{e_{ij}} \{\mathbf{k} \nabla \phi_k^i \cdot n\} \phi_l^j + \sigma [\phi_k^i] \phi_l^j d\Gamma & j \in \mathcal{N}_i \\ 0 & \text{otherwise} \end{cases}, \qquad (8c)$$

Note that the matrices $\mathbf{A}_i$ and $\mathbf{B}_{ij}$ depend on $\rho$, $c_p$ and $\mathbf{k}$, respectively, and hence can be a non-linear function of $\mathbf{u}$. Since the trial functions are orthogonal, if $\rho c_p$ is constant within an element, $\mathbf{A}_i$ is diagonal; otherwise, $\mathbf{A}_i$ is symmetric and positive definite, as $\rho c_p > 0$.

For compactness, the element-wise model Eq. (7) is also written in a matrix form

$$\mathbf{A}(\mathbf{u}) \dot{\mathbf{u}} = \mathbf{B}(\mathbf{u}) \mathbf{u} + \mathbf{f}(t), \qquad (9)$$

where $\mathbf{u} = [\mathbf{u}_1^\top, \mathbf{u}_2^\top, \cdots, \mathbf{u}_M^\top]^\top \in \mathbb{R}^{MP}$ includes all the DG variables, $\mathbf{f} = [\mathbf{f}_1^\top, \mathbf{f}_2^\top, \cdots, \mathbf{f}_M^\top]^\top \in \mathbb{R}^{MP}$, $\mathbf{A}$ is a matrix of $M$ diagonal blocks whose $i$th block is $\mathbf{A}_i$, and $\mathbf{B}$ is a matrix of $M \times M$ blocks whose $(i, j)$th block is

$$\mathbf{B}_{ij} = \begin{cases} \mathbf{B}_i + \sum_{l \in \mathcal{N}_i \cup \{T\}} \mathbf{B}_{il}^i & i = j \\ \mathbf{B}_{ij}^j & i \neq j \end{cases} \quad \text{for } i, j = 1, 2, \cdots, M. \qquad (10)$$

The dependency of $\mathbf{A}$ and $\mathbf{B}$ on $\mathbf{u}$ is explicitly noted in Eq. (9), which is the source of non-linearity in the current TPS problem.

## C. Reduced-Physics Model: Lumped Capacitance Model

The lumped capacitance model (LCM) is a classical physics-based low-order model for predicting the temporal variation of average temperature in multiple connected components [35]. The LCM is usually derived at the component level from a point view of energy conservation. The following assumptions are employed: (1) component $i$ has a uniform temperature $\bar{u}_i$; (2) between the neighboring components $i$ and $j$ the heat flux is,

$$q_{ij} = \frac{\bar{u}_j - \bar{u}_i}{R_{ij}}, \quad \text{for } i, j \in \mathcal{N}_i, \tag{11}$$

where $R_{ij}$ is the so-called thermal resistance. Empirically, for a component of isotropic heat conductivity $k$, length $L$ and cross-section area $A$, the thermal resistance is $R = L/kA$. Between components $i$ and $j$, define $R_{ij} = R_i + R_j$. In addition, the heat flux due to Dirichlet boundary condition is computed as $q_{iT} = (T_b - \bar{u}_i)/R_i$.

At component $i$, the dynamics of LCM is

$$\int_{\Omega_i} \rho c_p \dot{\bar{u}}_i d\Omega = \left( \sum_{j \in \mathcal{N}_i} \int_{e_{ij}} \frac{\bar{u}_j - \bar{u}_i}{R_{ij}} d\Gamma \right) + \int_{e_{iq}} q_b d\Gamma + \int_{e_{iT}} \frac{T_b - \bar{u}_i}{R_i} d\Gamma \tag{12a}$$

$$\Rightarrow \quad \bar{A}_i \dot{\bar{u}}_i = \left( \sum_{j \in \mathcal{N}_i} \frac{|e_{ij}|}{R_{ij}} (\bar{u}_j - \bar{u}_i) \right) + |e_{iq}| \bar{q}_i + \frac{|e_{iT}|}{R_i} (\bar{T}_i - \bar{u}_i) \tag{12b}$$

$$= \sum_{j \in \mathcal{N}_i} \left( -\frac{|e_{ij}|}{R_{ij}} \bar{u}_i + \frac{|e_{ij}|}{R_{ij}} \bar{u}_j \right) + \left( -\frac{|e_{iT}|}{R_i} \bar{u}_i \right) + \left( |e_{iq}| \bar{q}_i + \frac{|e_{iT}|}{R_i} \bar{T}_i \right) \tag{12c}$$

$$\equiv \sum_{j \in \mathcal{N}_i \cup \{T\}} \left( \bar{B}^i_{ij} \bar{u}_i + \bar{B}^j_{ij} \bar{u}_j \right) + \bar{f}_i, \tag{12d}$$

where in Eq. (12b) $|e|$ denotes the length ($d = 2$) or area ($d = 3$) of a component boundary $e$, and

$$\bar{A}_i = \int_{\Omega_i} \rho c_p d\Omega, \quad \bar{q}_i = \frac{1}{|e_{iq}|} \int_{e_{iq}} q_b d\Gamma, \quad \bar{T}_i = \frac{1}{|e_{iT}|} \int_{e_{iT}} T_b d\Gamma. \tag{13}$$

For compactness, the complete LCM is written in a matrix form,

$$\bar{\mathbf{A}}(\bar{\mathbf{u}})\dot{\bar{\mathbf{u}}} = \bar{\mathbf{B}}(\bar{\mathbf{u}})\bar{\mathbf{u}} + \bar{\mathbf{f}}(t), \tag{14}$$

where $\bar{\mathbf{u}} = [\bar{u}_1, \bar{u}_2, \cdots, \bar{u}_N]^\top \in \mathbb{R}^N$ includes average temperatures of all components, $\bar{\mathbf{f}} = [\bar{\mathbf{f}}_1^\top, \bar{\mathbf{f}}_2^\top, \cdots, \bar{\mathbf{f}}_M^\top]^\top \in \mathbb{R}^N$, $\bar{\mathbf{A}} \in \mathbb{R}^{N \times N}$ is a diagonal matrix whose $i$th element is $\bar{A}_i$, and for $\bar{\mathbf{B}} \in \mathbb{R}^{N \times N}$ the $(i, j)$th element is

$$\bar{B}_{ij} = \begin{cases} \sum_{j \in \mathcal{N}_i \cup \{T\}} \bar{B}^i_{ij} & i = j \\ \bar{B}^j_{ij} & i \neq j \end{cases} \quad \text{for } i, j = 1, 2, \cdots, N. \tag{15}$$

Similar to the DG model, the dependency of $\bar{\mathbf{A}}$ and $\bar{\mathbf{B}}$ on $\bar{\mathbf{u}}$ is explicitly noted in Eq. (14); this dependency partially captures the non-linearity in the TPS problem.

**D. Summary of Modeling Approaches**

The FOM (i.e., DG model) and RPM (i.e., LCM) are two different but mathematically connected solution strategies. On one hand, the FOM is the most accurate but computationally expensive to evaluate due to the fine mesh discretizations that translate to possibly millions of state variables. On the other hand, the RPM considers only the average temperature of the material as the state variables, considerably reducing the computational cost, but sacrificing the fidelity of the temperature prediction. Thus, neither the FOM nor the RPM is a universal approach for real-world analysis, design, and optimization tasks, where thousands of high-fidelity model evaluations may be necessary. This issue motivates the development of the PIROM that can achieve the fidelity of FOM at a computational cost close to the RPM, while maintaining the generalizability to model parameters.

# III. Physics-Infused Reduced-Order Modeling

The formulation of PIROM for TPS starts by mathematically connecting the full-order model, i.e., DG-FEM, and the reduced-physics model, i.e., LCM, via a coarse-graining procedure. This procedure pinpoints the missing dynamics in LCM when compared to DG-FEM. Subsequently, the Mori-Zwanzig formalism is employed to determine the model form for the missing dynamics in PIROM. Lastly, the data-driven identification of the missing dynamics in PIROM is presented.

**A. Deriving the Reduced-Physics Model via Coarse-Graining**

The LCM in Eq. (14) not only resembles the functional form of the DG model in Eq. (9), but can also be viewed as a special case of the latter, where the mesh partition is extremely coarse, and the trial and test functions are piece-wise constants. For example, consider the case where each component $\Omega_i$ is treated as one single element, and each element employs only one constant basis function $\phi_1^1 = 1$. The element-wise DG model in Eq. (7) simplifies into a scalar ODE, where,

$$\mathbf{A}_i = \bar{A}_i, \quad \mathbf{B}_i = 0, \quad \mathbf{B}_{ij}^i = -\sigma|e_{ij}|, \quad \mathbf{B}_{ij}^j = \sigma|e_{ij}|, \quad \mathbf{f}_i = |e_{iq}|\bar{q}_i + \sigma|e_{iT}|\bar{T}_i. \qquad (16)$$

Clearly, the LCM is a coarse zeroth-order DG model with the inverse of thermal resistance chosen as the element-wise penalty factors. Or conversely, the DG model is a refined version of LCM via $hp$-adaptation.

More precisely, the LCM can be obtained from a full-order DG model on a fine mesh via a projection process, or, coarse-graining. This process constrains the trial function space of a full-order DG model to a subset of piece-wise constants, so that the variables **u**, and the coefficient matrices **A** and **B**, and the vector **f** are all approximated using a single state associated to the average temperature. The details of the projection is described as follows.

*1. Coarse-graining of states*

Specifically, consider a DG model as in Eq. (9) for $M$ elements and a LCM as in Eq. (14) for $N$ components; clearly $M \gg N$. Let $\mathcal{V}_j = \{i | E_i \in \Omega_j\}$ be the indices of elements belonging to the $j$-th component, so $E_i \in \Omega_j$ for $i \in \mathcal{V}_j$; number of elements in $\Omega_j$ is denoted as $|\mathcal{V}_j|$. The average temperature in $\Omega_j$ is,

$$\bar{u}_j = \frac{1}{|\Omega_j|} \sum_{i \in \mathcal{V}_j} \int_{E_i} \boldsymbol{\phi}_i^\top \mathbf{u}_i d\Omega \equiv \frac{1}{|\Omega_j|} \sum_{i \in \mathcal{V}_j} |E_i| \boldsymbol{\varphi}_i^{j\top} \mathbf{u}_i, \quad j = 1, 2, \cdots, N. \tag{17}$$

Using the orthogonal basis functions chosen earlier, $\boldsymbol{\varphi}_i^j = [1, 0, \ldots, 0]^\top \in \mathbb{R}^P$. Conversely, given average temperatures of the $N$ components, $\bar{\mathbf{u}}$, the states of an arbitrary element $E_i$ is written as,

$$\mathbf{u}_i = \sum_{k=1}^{N} \boldsymbol{\varphi}_i^k \bar{u}_k + \delta \mathbf{u}_i, \tag{18}$$

where $\boldsymbol{\varphi}_i^k = 0$ if $i \notin \mathcal{V}_k$, and $\delta \mathbf{u}_i$ is the deviation from the average temperature and satisfies the orthogonality condition $\boldsymbol{\varphi}_i^{k\top} \delta \mathbf{u}_i = 0$ for all $k$.

Equations (17) and (18) can be written in matrix forms as, respectively,

$$\bar{\mathbf{u}} = \boldsymbol{\Phi}^+ \mathbf{u}, \quad \mathbf{u} = \boldsymbol{\Phi} \bar{\mathbf{u}} + \delta \mathbf{u}, \tag{19}$$

where $\boldsymbol{\Phi} \in \mathbb{R}^{MP \times N}$ is a matrix of $M \times N$ blocks, with the $(i, j)$th block as $\boldsymbol{\varphi}_i^j$, $\boldsymbol{\Phi}^+ \in \mathbb{R}^{N \times MP}$ is the left inverse of $\boldsymbol{\Phi}$, with the $(j, i)$th block as

$$\boldsymbol{\varphi}_i^{j+} = \frac{|E_i|}{|\Omega_j|} \boldsymbol{\varphi}_i^{j\top},$$

and $\delta \mathbf{u}$ is the collection of deviations. By their definitions, $\boldsymbol{\Phi}^+ \boldsymbol{\Phi} = \mathbf{I}$ and $\boldsymbol{\Phi}^+ \delta \mathbf{u} = 0$.

*2. Coarse-graining of dynamics*

Next, consider a function of states in the form of $\mathbf{M}(\mathbf{u})\mathbf{g}(\mathbf{u})$, where $\mathbf{g} : \mathbb{R}^{MP} \mapsto \mathbb{R}^{MP}$ is a vector-valued function and $\mathbf{M} : \mathbb{R}^{MP} \mapsto \mathbb{R}^{p \times MP}$ is a matrix-valued function with an arbitrary dimension $p$. Define projection matrix $\mathbf{P} = \boldsymbol{\Phi}\boldsymbol{\Phi}^+$ and projection operator $\mathcal{P}$ as

$$\mathcal{P}[\mathbf{M}(\mathbf{u})\mathbf{g}(\mathbf{u})] = \mathbf{M}(\mathbf{P}\mathbf{u})\mathbf{P}\mathbf{g}(\mathbf{P}\mathbf{u}) = \mathbf{M}(\boldsymbol{\Phi}\bar{\mathbf{u}})\mathbf{P}\mathbf{g}(\boldsymbol{\Phi}\bar{\mathbf{u}}), \tag{20}$$

so that the resulting function depends only on the average temperatures $\bar{\mathbf{u}}$. Correspondingly the residual operator is $\mathcal{Q} = \mathcal{I} - \mathcal{P}$, and

$$\mathcal{Q}[\mathbf{M}(\mathbf{u})\mathbf{g}(\mathbf{u})] = \mathbf{M}(\mathbf{u})\mathbf{g}(\mathbf{u}) - \mathcal{P}[\mathbf{M}(\mathbf{u})\mathbf{g}(\mathbf{u})].$$

Subsequently, the operators defined above are applied to coarse-grain the dynamics. First, write the DG model, Eq. (9), as

$$\dot{\mathbf{u}} = \mathbf{A}(\mathbf{u})^{-1}\mathbf{B}(\mathbf{u})\mathbf{u} + \mathbf{A}(\mathbf{u})^{-1}\mathbf{f}(t) \equiv \mathbf{r}(\mathbf{u}, t),$$

and multiplying both sides by $\mathbf{\Phi}^+$ produces

$$\mathbf{\Phi}^+ \dot{\mathbf{u}} = \mathbf{\Phi}^+(\mathbf{\Phi}\dot{\bar{\mathbf{u}}} + \delta\dot{\mathbf{u}}) = \dot{\bar{\mathbf{u}}} = \mathbf{\Phi}^+ \mathbf{r}(\mathbf{u}, t).$$

Applying the projection and residual operators,

$$\dot{\bar{\mathbf{u}}} = \mathcal{P}\left[\mathbf{\Phi}^+ \mathbf{r}(\mathbf{u}, t)\right] + \mathcal{Q}\left[\mathbf{\Phi}^+ \mathbf{r}(\mathbf{u}, t)\right] \equiv \mathbf{r}^{(1)}(\bar{\mathbf{u}}, t) + \mathbf{r}^{(2)}(\mathbf{u}, t), \tag{21}$$

where $\mathbf{r}^{(1)}(\bar{\mathbf{u}}, t)$ is referred to as the resolved dynamics, that depends only on $\bar{\mathbf{u}}$, and $\mathbf{r}^{(2)}(\mathbf{u}, t)$ is referred to as the residual dynamics.

While the detailed derivations and analysis of the $\mathbf{r}^{(1)}$ and $\mathbf{r}^{(2)}$ terms are provided in Appendix A, as well as an illustrative one-dimensional example, the key results are stated below.

First, the resolved dynamics is exactly the LCM. Specifically, using the notation in Eq. (14),

$$\mathbf{r}^{(1)}(\bar{\mathbf{u}}, t) = \bar{\mathbf{A}}(\bar{\mathbf{u}})^{-1}\bar{\mathbf{B}}(\bar{\mathbf{u}})\bar{\mathbf{u}} + \bar{\mathbf{A}}(\bar{\mathbf{u}})^{-1}\bar{\mathbf{f}}(t), \tag{22}$$

where the following relations hold,

$$\bar{\mathbf{A}}(\bar{\mathbf{u}}) = \mathbf{W}\left(\mathbf{\Phi}^+ \mathbf{A}(\mathbf{\Phi}\bar{\mathbf{u}})^{-1}\mathbf{\Phi}\right)^{-1}, \quad \bar{\mathbf{B}}(\bar{\mathbf{u}}) = \mathbf{W}\mathbf{\Phi}^+ \mathbf{B}(\mathbf{\Phi}\bar{\mathbf{u}})\mathbf{\Phi}, \quad \bar{\mathbf{f}} = \mathbf{W}\mathbf{\Phi}^+ \mathbf{f}. \tag{23}$$

with a diagonal weight matrix $\mathbf{W} \in \mathbb{R}^{M \times M}$, such that the $i$th diagonal element $[\mathbf{W}]_i = |\mathcal{V}_k|$ if $i \in \mathcal{V}_k$.

Second, the examination of the $\mathbf{r}^{(2)}$ term shows the physical sources of the missing dynamics in LCM. Take the forcing term in $\mathbf{r}^{(2)}$ as an illustration,

$$\left\|\mathbf{\Phi}^+ \mathbf{A}^{-1}\mathbf{f} - \bar{\mathbf{A}}^{-1}\bar{\mathbf{f}}\right\| \leq \left\|\mathbf{\Phi}^+\right\| \underbrace{\left\|\mathbf{A}(\mathbf{u})^{-1} - \mathbf{A}(\mathbf{\Phi}\bar{\mathbf{u}})^{-1}\right\|}_{\#1} \|\mathbf{f}\| + \left\|\mathbf{\Phi}^+ \mathbf{A}(\mathbf{\Phi}\bar{\mathbf{u}})^{-1}\right\| \underbrace{\left\|\mathbf{f} - \mathbf{\Phi}\bar{\mathbf{f}}\right\|}_{\#2}, \tag{24}$$

where the norms #1 and #2 are the errors due to the non-uniformities of material properties and heat flux, respectively.

In sum, the above results not only show that the LCM is a result of coarse-graining of the full-order DG model, but also reveal the discrepancies between the LCM and the DG model. In the subsequent section, the discrepancies will be corrected to produce the proposed PIROM.

## B. Formulation of Reduced-Order Model

The Mori-Zwanzig (MZ) formalism is an operator-projection technique used to derive ROMs for high-dimensional dynamical systems, especially in statistical mechanics and fluid dynamics [30–32, 36]. It provides an exact reformulation of the full-order dynamics in terms of a subset of resolved states. The proposed ROM is subsequently developed based on such reformulation.

First, the results of the previous section show that the dynamics of the resolved variables $\bar{\mathbf{u}}$ can be decomposed in the form of Eq. (21), where one component $\mathbf{r}^{(1)}$ depends only on the resolved variables and the other component $\mathbf{r}^{(2)}$ is orthogonal to $\mathbf{r}^{(1)}$, in the sense of $\mathcal{P}\mathbf{r}^{(2)} = 0$. In this case, the MZ formalism can be invoked to express the dynamics of $\bar{\mathbf{u}}$ entirely in terms of $\bar{\mathbf{u}}$ alone as an integro-differential equation [32],

$$\dot{\bar{\mathbf{u}}} = \mathbf{r}^{(1)}(\bar{\mathbf{u}}, t) + \int_0^t \tilde{\kappa}(t, s, \bar{\mathbf{u}})\, ds, \tag{25}$$

where the first term on the right is the Markovian term, the integral is referred to as the non-Markovian term, and $\tilde{\kappa}$ is the memory kernel. The integral accounts for the memory effect, i.e., the impact of past resolved states on the current states through their interactions with unresolved states.

Next, to further inform the subsequent derivation of the ROM, the kernel $\tilde{\kappa}$ is examined via a leading-order expansion, based on prior work [29]; this can be viewed as an analog of zeroth-order hold in linear system theory with a sufficiently small step size. In this case, the integrand in the non-Markovian term is approximated as,

$$\tilde{\kappa}(t, s, \bar{\mathbf{u}}) \approx \mathbf{r}^{(1)}(\bar{\mathbf{u}}, t) \cdot \nabla_{\bar{\mathbf{u}}} \mathbf{r}^{(2)}(\boldsymbol{\Phi}\bar{\mathbf{u}}, t). \tag{26}$$

Here note that the terms in $\mathbf{r}^{(1)}$ have a common factor of $\bar{\mathbf{A}}^{-1}$; this motivates the following heuristic modification of the model form in Eq. (25),

$$\dot{\bar{\mathbf{u}}} = \mathbf{r}^{(1)}(\bar{\mathbf{u}}, t) + \bar{\mathbf{A}}^{-1}(\bar{\mathbf{u}}) \int_0^t \kappa(t, s, \bar{\mathbf{u}})\, ds$$
$$\Rightarrow \quad \bar{\mathbf{A}}(\bar{\mathbf{u}})\dot{\bar{\mathbf{u}}} = \bar{\mathbf{B}}(\bar{\mathbf{u}})\bar{\mathbf{u}} + \bar{\mathbf{f}}(t) + \int_0^t \kappa(t, s, \bar{\mathbf{u}})\, ds, \tag{27}$$

where the original kernel $\tilde{\kappa}$ is effectively normalized by $\bar{\mathbf{A}}^{-1}$. Intuitively, such choice of kernel $\kappa$ reduces its dependency on the averaged material properties, and simplifies the subsequent design of model form.

Subsequently, hidden states are introduced to make the non-Markovian component Markovian [29]. In this manner, Eq. (27) is converted to a pure state-space model, with the functional form of LCM retained; since LCM is a physics-based model, then it encodes the physical information and retains explicit parametric dependence of the problem. Consider the representation of the kernel as a finite sum of simpler functions, e.g., exponentials,

$$\kappa(t, s, \bar{\mathbf{u}}) \approx \sum_{j=1}^{m} e^{-\lambda_j(t-s)} \mathbf{p}_j \left( \mathbf{q}_j^\top \bar{\mathbf{u}}(s) + \mathbf{r}_j^\top \bar{\mathbf{f}}(s) \right), \tag{28}$$

with suitable coefficients $\mathbf{p}_j, \mathbf{q}_j, \mathbf{r}_j \in \mathbb{R}^N$ and decay rates $\lambda_j > 0$, that need to be identified from data.

Define the hidden states as,

$$\beta_j = \int_0^t e^{-\lambda_j(t-s)} \left( \mathbf{q}_j^\top \bar{\mathbf{u}}(s) + \mathbf{r}_j^\top \bar{\mathbf{f}}(s) \right) ds, \tag{29}$$

then through its differentiation with respect to time,

$$\dot{\beta}_j = -\lambda_j \beta_j + \mathbf{q}_j^\top \bar{\mathbf{u}} + \mathbf{r}_j^\top \bar{\mathbf{f}}, \tag{30}$$

and the memory kernel becomes,

$$\int_0^t \kappa(t, s, \bar{\mathbf{u}}) \, ds = \sum_{j=1}^{m} \mathbf{p}_j \beta_j. \tag{31}$$

Then, Eq. (27) is recast as an extended Markovian system,

$$\bar{\mathbf{A}}(\bar{\mathbf{u}}) \dot{\bar{\mathbf{u}}} = \bar{\mathbf{B}}(\bar{\mathbf{u}}) \bar{\mathbf{u}} + \bar{\mathbf{f}}(t) + \mathbf{P}\boldsymbol{\beta} \tag{32a}$$

$$\dot{\boldsymbol{\beta}} = \mathbf{Q}\bar{\mathbf{u}} - \Lambda\boldsymbol{\beta} + \mathbf{R}\bar{\mathbf{f}}(t), \tag{32b}$$

where $\Lambda = \text{diag}(\lambda_1, \lambda_2, \cdots, \lambda_m)$, $\mathbf{P} = [\mathbf{p}_1, \mathbf{p}_2, \cdots, \mathbf{p}_m] \in \mathbb{R}^{N \times m}$, $\mathbf{Q} = [\mathbf{q}_1, \mathbf{q}_2, \cdots, \mathbf{q}_m] \in \mathbb{R}^{m \times N}$, and $\mathbf{R} = [\mathbf{r}_1, \mathbf{r}_2, \cdots, \mathbf{r}_m] \in \mathbb{R}^{m \times N}$. Since the hidden states $\boldsymbol{\beta}$ serve as the memory, their initial conditions are always $\boldsymbol{\beta}(0) = 0$, i.e., no memory at the beginning.

Lastly, denote the collection of resolved and hidden states as $\mathbf{y} = [\bar{\mathbf{u}}, \boldsymbol{\beta}] \in \mathbb{R}^{n_y}$, $n_y = N + m$, Eq. (32) is written more generally as

$$\tilde{\mathbf{A}}\dot{\mathbf{y}} = \mathbf{B}\mathbf{y} + \mathbf{C}\bar{\mathbf{f}} \tag{33a}$$

$$\mathbf{z} = \mathbf{D}\mathbf{y}, \tag{33b}$$

where

$$\tilde{\mathbf{A}} = \begin{bmatrix} \bar{\mathbf{A}}(\bar{\mathbf{u}}) & \mathbf{O} \\ \mathbf{O} & \mathbf{I} \end{bmatrix} \in \mathbb{R}^{n_y \times n_y}, \ \mathbf{B} = \begin{bmatrix} \bar{\mathbf{B}}(\bar{\mathbf{u}}) & \mathbf{P} \\ \mathbf{Q} & \mathbf{E} \end{bmatrix} \in \mathbb{R}^{n_y \times n_y}, \ \mathbf{C} = \begin{bmatrix} \mathbf{I} \\ \mathbf{R} \end{bmatrix} \in \mathbb{R}^{n_y \times N}, \ \mathbf{D} \in \mathbb{R}^{n_z \times n_y}.$$

In Eq. (33a), the terms $\bar{\mathbf{A}}, \bar{\mathbf{B}},$ and $\bar{\mathbf{f}}$ are the LCM terms, which explicitly depend on the temperature-dependent material properties. The collection of matrices $\Theta = \{\mathbf{P}, \mathbf{Q}, \mathbf{E}, \mathbf{R}, \mathbf{D}\} \in \mathbb{R}^{n_\theta}$ are learnable parameters to capture the memory effects. Particularly note that $\mathbf{\Lambda}$ is replaced by a fully-populated matrix $\mathbf{E}$ for potentially better generalizability and easier training. In Eq. (33b), $\mathbf{D}$ is a fully-populated matrix that extracts the observables from the PIROM states $\mathbf{y}$.

### C. Learning the Reduced-Order Model from Data

After the design of model structure of PIROM, the last step is to learn the unknown dynamics from observed data. For the ease of presentation, consider the following compact form of the PIROM, Eq. (33),

$$\mathcal{F}\left(\dot{\mathbf{y}}, \mathbf{y}; \boldsymbol{\xi}, \boldsymbol{\Theta}\right) = 0, \tag{34}$$

where $\boldsymbol{\xi}$ defines the operating condition, such as the BC's, i.e., the heat flux, and material properties. Consider a dataset of $N_s$ high-fidelity trajectories of observables over a time interval $[t_0, t_f]$,

$$\mathcal{D}_{HF} = \left\{ \left( t_i, \mathbf{z}_{HF}^{(l)}(t_i), \boldsymbol{\xi}^{(l)} \right) \right\}_{l=1}^{N_s}, \quad i = 1, 2, \ldots, N_p. \tag{35}$$

The learning problem is formulated as the following trajectory minimization problem over all observables,

$$\min_{\boldsymbol{\Theta}} \quad \mathcal{J}(\boldsymbol{\Theta}; \mathcal{D}_{HF}) = \sum_{i=1}^{N_s} \int_{t_0}^{t_f} \ell\left(\mathbf{z}^{(l)}, \mathbf{z}_{HF}^{(l)}\right) dt \tag{36a}$$

$$\text{s.t.} \quad 0 = \mathcal{F}\left(\dot{\mathbf{y}}^{(l)}, \mathbf{y}^{(l)}; \boldsymbol{\xi}^{(l)}, \boldsymbol{\Theta}\right) \tag{36b}$$

$$\mathbf{z}^{(l)} = \mathbf{D}\mathbf{y}^{(l)}, \quad l = 1, 2, \cdots, N_s, \tag{36c}$$

where the objective is to minimize the discrepancy between the high-fidelity and PIROM temperature observable predictions for the $l$-th trajectory, and the integrand is

$$\ell\left(\mathbf{z}^{(l)}, \mathbf{z}_{HF}^{(l)}\right) = \left\| \mathbf{z}^{(l)} - \mathbf{z}_{HF}^{(l)} \right\|_2^2. \tag{37}$$

The learning problem is solved using a gradient-based optimization algorithm following the concept of neural ODE [6]. This algorithm requires the computation of $\nabla_{\boldsymbol{\Theta}} \mathcal{J}$, which requires the sensitivity of the differential constraints $\frac{d\mathbf{y}^{(l)}}{d\boldsymbol{\Theta}}$. The sensitivity may be computed by solving the following first-order necessary conditions for optimality (FONC) with respect to the $\Theta$ variables [37],

$$\mathcal{F}(\mathbf{y}, \dot{\mathbf{y}}; \boldsymbol{\xi}, \Theta) = 0 \tag{38a}$$

$$\frac{\partial \ell}{\partial \mathbf{z}} \frac{\partial \mathbf{z}}{\partial \mathbf{y}} + \lambda^\top \frac{\partial \mathcal{F}}{\partial \mathbf{y}} - \frac{d}{dt}\left(\lambda^\top \frac{\partial \mathcal{F}}{\partial \dot{\mathbf{y}}}\right) = 0 \tag{38b}$$

$$\lambda(t_f)^\top \frac{\partial \mathcal{F}}{\partial \dot{\mathbf{y}}(t_f)} = 0, \tag{38c}$$

where $\lambda$ are the adjoint variables, representing the sensitivity of the objective function with respect to the dynamical constraints. Equations (38a)-(38b) are the state and adjoint dynamics, respectively, and Eq. (38c) specifies the boundary condition for the adjoint variables based on the terminal cost; for this problem $\lambda(t_f) = 0$. Once Eq. (38) is solved, the gradient is computed as

$$\nabla_\Theta \mathcal{J} = \frac{1}{N_s} \sum_{l=1}^{N_s} \int_{t_0}^{t_f} \left(\frac{\partial \ell}{\partial \Theta} + \left(\lambda^{(l)}\right)^T \frac{\partial \mathcal{F}}{\partial \Theta}\right) dt. \tag{39}$$

In practice, directly learning the PIROM for the desired (long) time horizon of $t_f$ at a low tolerance $\epsilon^s$ may suffer from numerical instability in the form of, e.g., slow convergence or divergence during training. To improve the training efficiency of the PIROM over the time horizon of interest, a tolerance-sweeping training approach is adopted to gradually decrease training tolerances and increase the time horizon, as summarized in Alg. 1. The algorithm initializes with a decreasing set of tolerances $\mathbb{E} = \{\epsilon^i\}_{i=1}^s$, an increasing set of time horizons $\mathbb{T} = \{t_f^j\}_{j=1}^p$ with $t_f^p = t_f$, and the maximum number of iterations for each time horizon $N_{\max}$. The algorithm starts with a relatively easy optimization, that minimizes the trajectory error over a short horizon $\left[t_0, t_f^1\right]$ until either the tolerance $\epsilon^1$ or the maximum number of iterations is reached. Subsequently, at the $j$th stage, $j = 2, 3, \cdots, p$, the algorithm minimizes the trajectory error with the time horizon extended to $\left[t_0, t_f^j\right]$. Subsequently, after the $p$ stages of training is done for tolerance $\epsilon^1$, these stages are repeated for tolerances $\epsilon^i$, for $i = 2, 3, \cdots, s$, until the desired convergence is achieved.

## IV. Application to Thermal Protection Systems

In this section, the proposed PIROM approach is applied to a typical TPS problem, and is compared against representative data-driven reduced-order models for TPS, including the Operator Inference (OpInf) [23], that selects the model form based on physical knowledge, and the neural ordinary differential equation (NODE) [6], that uses a generic neural network to represent the dynamics. The model benchmark quantifies: (1) the generalizability of the models to unseen load conditions and material properties, and (2) the computational cost associated with model training and evaluation. The results show PIROM to be a promising candidate for the solution of the impossible trinity of modeling.

**Algorithm 1:** Tolerance-sweeping training algorithm for the PIROM.

1 **Input:** $\mathbb{E} = \{\epsilon^i\}_{i=1}^s$, $\mathbb{T} = \{t_f^j\}_{j=1}^p$, $N_{\max}$, and $\Theta_0$
2 **Require:** $\epsilon^1 > \epsilon^2 > \cdots > \epsilon^s$, $t_f^1 < t_f^2 < \cdots < t_f^p$
3 **for** $i = 1 \ldots s$ **do**
4    **for** $j = 1 \ldots p$ **do**
5       **Set:** $k = 0$, $\mathcal{J}_k = \infty$
6       **while** $\mathcal{J}_k \geq \epsilon^i$ and $k \leq N_{\max}$ **do**
7          Select $n$ random cases from dataset $\mathcal{D}$: $\Xi = \{\boldsymbol{\xi}^{(1)}, \boldsymbol{\xi}^{(2)}, \cdots, \boldsymbol{\xi}^{(n)}\}$
8          **for** $\boldsymbol{\xi} \in \Xi$ **do**
9             Solve Eq. (38) given $\boldsymbol{\xi}^{(l)}$, $\Theta_k$, $[t_0, t_f^j]$ to obtain $\boldsymbol{\lambda}^{(l)}$ using `torchdiffeq`
10         Numerically integrate Eq. (39) given $\boldsymbol{\lambda}^{(l)}$ to obtain $\nabla_\Theta \mathcal{J}^{(l)}$
11          $\nabla_\Theta \mathcal{J} \leftarrow \frac{1}{n} \sum_{l=1}^n \nabla_\Theta \mathcal{J}^{(l)}$
12          $\Theta_{k+1} \leftarrow \text{ADAM}(\Theta_k, \nabla_\Theta \mathcal{J})$
13          $\mathcal{J}_k \leftarrow \mathcal{J}(\Theta_{k+1}; \mathcal{D}_{\text{HF}})$
14          $k = k + 1$
15       **Set:** $\Theta_0 = \Theta_k$
16 **Result:** *Optimized parameter* $\Theta_0$.

| Layer | Material | Density, $\left[\frac{\text{kg}}{\text{m}^3}\right]$ | Dimensions, [m] | | Boundary Conditions | | | |
|---|---|---|---|---|---|---|---|---|
| | | | Thickness $t$ | Width $w$ | Top | Left | Right | Bottom |
| 1 | Acusil | 256.3 | 0.01 | 0.04 | N | - | A | - |
| 2 | Tungsten | 18740 | 0.02 | 0.01 | N | A | - | A |
| 3 | Titanium | 5199 | 0.01 | 0.04 | - | - | A | A |

**Table 1** The TPS configuration. For BCs: N: Neumann; A: Adiabatic.

### A. Problem Definition

Consider the two-dimensional TPS configuration in Fig. 3 with materials, dimensions, and BCs listed in Table 1. Such configuration is representative of the TPS used for the initial concept 3.X vehicle in past studies [38], and involves three layers of different materials: Acusil, Tungsten, and Titanium. At the top surface, the Acusil and Tungsten layers are directly subjected to strong time-varying and non-uniform heating, while the rest of the surfaces are assumed to be adiabatic. The source of non-linearity in the problem comes from the temperature-dependent material properties of the layers, which are shown in black lines in Fig. 3. Only the material densities are temperature independent, which have been listed in Table 1. Moreover, as shown in Fig. 3, perfect thermocouple devices are placed at the centroids of each layer for the collection of temperature signals that are used in the following sections for training and testing the ROMs.

*1. Parametrization of Boundary Conditions and Material Properties*

The operating condition of the TPS is determined by the boundary conditions, i.e., the heat flux, and material properties.

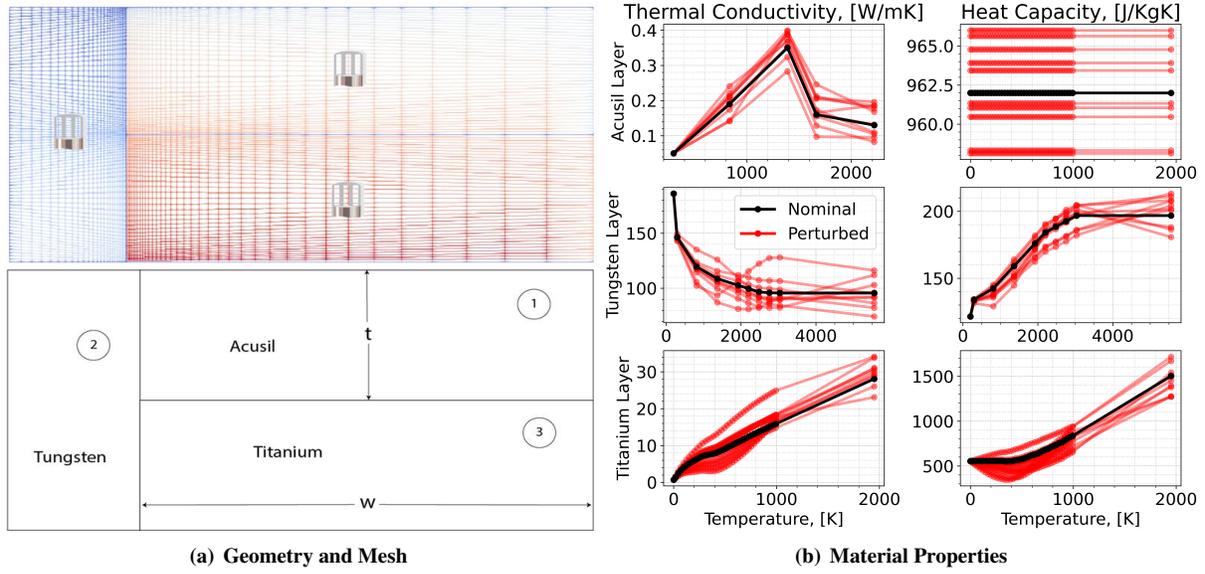

(a) Geometry and Mesh  (b) Material Properties

Fig. 3  Geometry and material properties of the TPS.

The heat flux on the Neumann BC is parametrized using $\boldsymbol{\xi}_{BC} = \{\xi_0, \xi_1, \xi_2\}$, and defined as

$$q(x,t) = \frac{1}{\sqrt{800\pi}} \left(\xi_0 + \xi_1 \left(\frac{x+a}{b}\right)\right) \left(1 + \xi_2 \exp\left(-\left(\frac{t-600}{400}\right)^2\right)\right), \qquad (40)$$

where $\xi_0$ controls the magnitude of the average heat flux, $\xi_1$ the slope of the distribution in space, and $\xi_2$ the rate of change of the magnitude with respect to time. When $\xi_1 = \xi_2 = 0$, the heat flux is uniform in space and constant over time.

The material properties are parametrized using $\boldsymbol{\xi}_M = \{\alpha_0, \alpha_1, \alpha_2\}$. Let $m$ represent either the thermal conductivity or specific heat capacity of a material, then the material properties are perturbed as in Eq. (41),

$$m^{(p)}(\bar{T}) = m^{(n)}(\bar{T}) + 0.5 \Delta m^{(n)} \left(\alpha_0 \bar{T} + \alpha_1 \sin\left(\alpha_2 \pi \bar{T}\right)\right), \qquad (41)$$

where $m^{(p)}$ and $m^{(n)}$ correspond to perturbed and nominal material properties, respectively. In Eq. (41), the temperature is normalized as $\bar{T} = \frac{T - T_{\min}}{T_{\max} - T_{\min}}$, where $T_{\min}$ and $T_{\max}$ are the minimum and maximum temperatures available in the material property datasets, and $\Delta m^{(n)} = \max(m^{(n)}) - \min(m^{(n)})$ is the range of the nominal material properties. The perturbations are defined using linear and sinusoidal functions that increase in magnitude as the temperature increases, as shown in Fig. 3. Increasing the perturbation with respect to temperature is intended to mimic the increasing uncertainty in material properties at high temperatures.

*2. Data Generation*

Full-order solutions of the TPS are computed using the FEM multi-mechanics module of the `Aria` package [39], where the mesh is shown in Fig. 3(a). The mesh consists of 5103 nodes with 1480, 2000, and 1480 elements assigned to the Acusil, Tungsten, and Titanium layers, respectively. The mesh is refined in regions where the temperature gradients are expected to be high, i.e., at the interfaces between the three layers. All solutions are computed up to $t_f$ = 1500 seconds from an uniform temperature of $T(x, t_0)$ = 300K. Given an operating condition $\boldsymbol{\xi}^{(l)} = [\boldsymbol{\xi}_{BC}^{(l)}, \boldsymbol{\xi}_{M}^{(l)}]$, a full-order solution consists of the collection of time-varying temperature fields $\left\{\left(t_k, \mathbf{u}_{HF}^{(l)}(t_k), \boldsymbol{\xi}^{(l)}\right)\right\}_{k=0}^{N_P}$, where $N_P$ is the number of time steps with a step size of $\Delta t$ = 1.5 seconds. The observable trajectories are representative of near-wall thermocouple sensing of hypersonic flows involving heat transfer. At each time instance $t_k$, a temperature reading is recorded in each layer using the thermocouples shown in Fig. 3(a), resulting in three temperature signals, i.e., the observations $\mathbf{z} \in \mathbf{R}^3$. Therefore, each full-order solution produces one trajectory of observables $\left\{\left(t_k, \mathbf{z}_{HF}^{(l)}(t_k), \boldsymbol{\xi}^{(l)}\right)\right\}_{k=0}^{N_P}$. The goal of the ROMs is to predict $\mathbf{z}_{HF}$ as accurately as possible.

Lastly, due to their specific formulations, PIROM and NODE are trained using only the observable trajectories (i.e., $\mathbf{z}_{HF}$), while OpInf is trained using the full-state data (i.e., $\mathbf{u}_{HF}$).

*3. Definition of Training and Testing Datasets*

The training and testing datasets are listed in Table 2, and are designed to: (1) minimize the information that a ROM can "see", and (2) to maximize the variability of test operating conditions to examine the ROM's generalization performance, respectively. The data samples in the parameter space are visualized in Fig. 4(a), and the corresponding trajectory data are shown in Fig. 4(b). The training dataset $\mathcal{D}_1$ includes only uniform constant heat flux ($\boldsymbol{\xi}_{BC} = [\xi_0, 0, 0]$) and nominal material properties ($\boldsymbol{\xi}_M = 0$), and is used to train all the ROMs.

Seven test datasets are considered in an increasing level of challenge. Dataset $\mathcal{D}_2$ includes the same parametrization as in $\mathcal{D}_1$, and serves as a basic benchmark of the ROMs to assess model accuracy. In this case, $\mathcal{D}_2$ is in-distribution; all models only need to interpolate within the training dataset and are expected to produce accurate predictions. The rest of test datasets are all out-of-distribution (OOD) and meant for testing the generalizability of the ROMs. Datasets $\{\mathcal{D}_i\}_{i=3}^{7}$ consider the same parametrization as in $\mathcal{D}_1$, but includes space-time varying heat fluxes that the ROMS have not seen during training; these case examine the generalizability of PIROM to unseen loading conditions. Specifically, $\mathcal{D}_3$ is randomly sampled over the three-dimensional parameter space of $\boldsymbol{\xi}$ and used for convergence study. For a systematic comparison of the ROMs, the datasets $\{\mathcal{D}_i\}_{i=4}^{7}$ consist of uniform grid samples in a $\xi_1 - \xi_2$ plane with 10 points per dimension; the values of $\xi_0$ are set to the mean, lower, and upper bounds for the distribution of $\xi_0$, respectively. Lastly, while all the previous datasets are generated using the nominal material properties, the dataset $\mathcal{D}_8$ considers space-time varying heat flux together with perturbations in the material properties displayed in Fig. 3(b). This is the most challenging case for the generalizability of the ROMs.

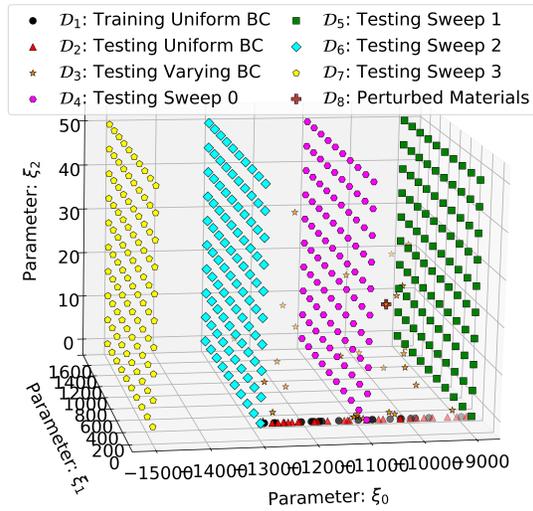

(a) **Parameter space of heat flux**

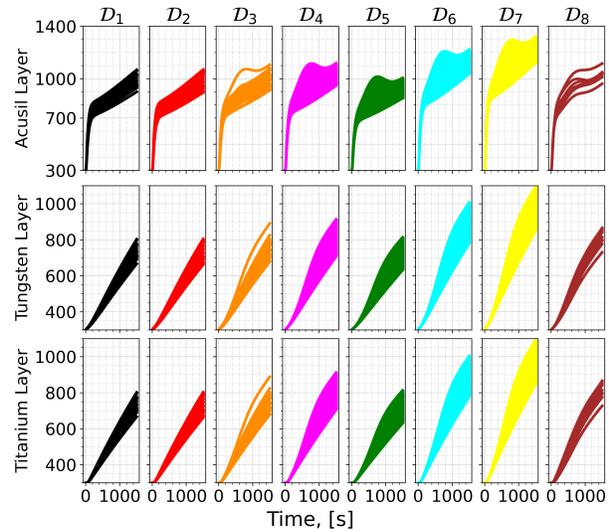

(b) **Trajectories of observed data**

**Fig. 4   Training and testing samples.**

| Dataset | | Models | $N_s$ | BC | $\xi_0 \times 10^3$ | $\xi_1 \times 10^3$ | $\xi_2$ | $\alpha_0$ | $\alpha_1$ | $\alpha_2$ |
|---|---|---|---|---|---|---|---|---|---|---|
| Train | $\mathcal{D}_1$ | A,B,C | 20 | Uniform | [9, 13] | 0 | 0 | 0 | 0 | 0 |
| Test | $\mathcal{D}_2$ | A,B,C | 20 | Uniform | [9, 13] | 0 | 0 | 0 | 0 | 0 |
| | $\mathcal{D}_3$ | A,B,C | 30 | Varying | [9, 13] | [0, 1.3] | [0, 40] | 0 | 0 | 0 |
| | $\mathcal{D}_4$ | A,B,C | 100 | Varying | 11 | [0, 1.5] | [0, 50] | 0 | 0 | 0 |
| | $\mathcal{D}_5$ | A,B,C | 100 | Varying | 9 | [0, 1.5] | [0, 50] | 0 | 0 | 0 |
| | $\mathcal{D}_6$ | A,B,C | 100 | Varying | 13 | [0, 1.5] | [0, 50] | 0 | 0 | 0 |
| | $\mathcal{D}_7$ | A,B,C | 100 | Varying | 15 | [0, 1.5] | [0, 50] | 0 | 0 | 0 |
| | $\mathcal{D}_8$ | A | 10 | Varying | 10.428 | 0.284 | 21 | $[-\frac{1}{2}, \frac{1}{2}]$ | $[-\frac{1}{2}, \frac{1}{2}]$ | $[-3, 3]$ |

**Table 2   Definition of training and testing datasets. A: PIROM; B: OpInf; C: NODE.**

## 4. Performance Metrics

The performance of the ROMs is evaluated by the metrics of prediction error and computational cost.

**Prediction error** Consider one trajectory of high-fidelity observed data $\left\{\left(t_k, \mathbf{z}_{HF}^{(l)}(t_k)\right)\right\}_{k=0}^{N_P}$ and ROM prediction $\left\{\left(t_k, \mathbf{z}^{(l)}(t_k)\right)\right\}_{k=0}^{N_P}$. The difference $e_i^{(l)}$ for the $i$th component of observation, denoted $z_i$, is quantified as,

$$e_i^{(l)} = \frac{1}{\Delta z_i^{(l)}} \sqrt{\frac{1}{N_P} \sum_{k=0}^{N_P} \left(z_{HF,i}^{(l)}(t_k) - z_i^{(l)}(t_k)\right)^2}, \quad \text{for } i = 1, 2, 3, \tag{42}$$

where

$$\Delta z_i^{(l)} = \max_{k=0,1,\cdots,N_P} \left(z_{HF,i}^{(l)}(t_k)\right) - \min_{k=0,1,\cdots,N_P} \left(z_{HF,i}^{(l)}(t_k)\right).$$

Subsequently, the prediction error of one trajectory is computed by a weighted sum based on the area of each layer, resulting in the normalized root mean square error (NRMSE) metric for one trajectory,

$$\text{NRMSE} = \frac{\sum_{i=1}^{3} |\Omega_i| e_i^{(l)}}{\sum_{i=1}^{3} |\Omega_i|} \times 100\%. \tag{43}$$

For one dataset, the NRMSE is defined to be the average of the NRMSEs of all the trajectories in the dataset.

**Computational cost** To ensure a fair comparison of the cost, all the computational costs discussed below are measured in serial computation, because the OpInf is usually solved by least squares in a serial mode. All the measurements are done on a single core on a AMD EPYC 9354P 32-Core CPU with 500 GB of RAM.

Denote $\mathcal{T}_\mathcal{M}^p(\mathcal{D})$ as the computational cost for a model $\mathcal{M}$ to predict the trajectories in a dataset $\mathcal{D}$. Here $\mathcal{M}$ can be "HF", denoting the full-order FEM solver, as well as the ROMs, i.e., "PIROM", "OpInf", and "NODE". Also denote $\mathcal{T}_\mathcal{M}^t(\mathcal{D})$ as the computational cost for a model $\mathcal{M}$ to train using a dataset $\mathcal{D}$. Specific metrics to consider for the computational cost are (1) the acceleration ratio $\mathcal{T}_{HF}^p/\mathcal{T}_\mathcal{M}^p$, which directly measures the acceleration obtained by the model $\mathcal{M}$ when compared to the full-order model, and (2) the overhead ratio $\mathcal{T}_\mathcal{M}^t/\mathcal{T}_{HF}^p$, which quantifies the overhead in creating the model $\mathcal{M}$.

**Summary of metrics** Using the metrics above, the NRMSE of in-distribution dataset (i.e., $\mathcal{D}_2$) quantifies the model accuracy, and the NRMSEs of OOD datasets (i.e., $\mathcal{D}_3 \sim \mathcal{D}_8$) quantify the model generalizability. The ratios of the computational costs quantify the computational efficiency of the models. An ideal ROM should achieve favorable values in all the metrics.

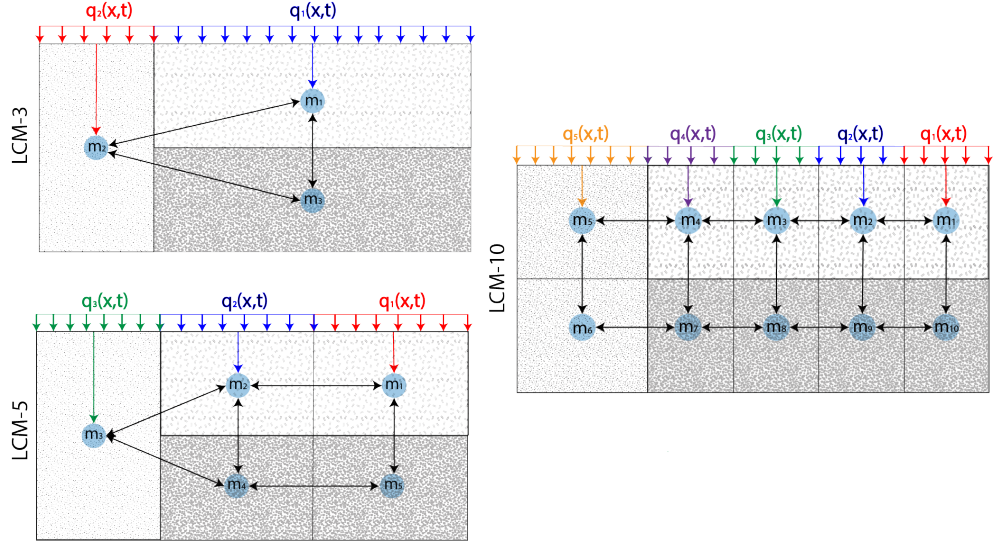

**Fig. 5    Lumped-mass representation for the three-layered TPS.**

## B. Implementation of the Physics-Infused Reduced-Order Model

The implementation details of the PIROM are provided next. The implementation details of OpInf and NODE are provided in Appendix B.

*1. Determining the Physics Component*

The physics component of the PIROM is based on the LCM, presented in Sec. II.C. To examine the trade-offs between model complexity and predictive performance, three LCM variants – LCM-3, LCM-5, and LCM-10 – are constructed, each discretizing the thermal response of the TPS into 3, 5, and 10 lumped masses, respectively. These configurations are illustrated in Fig. 5.

The distributed heat flux acting on the Acusil and Tungsten layers is represented by the uni-directional colored arrows feeding into the lumped masses. This distribution is divided by the number of components exposed to the hypersonic heat load. The bi-directional black arrows between the lumped masses represents the interacting heat transfer dynamics between neighboring components. The mathematical forms of LCM-3 and LCM-5 models are provided in Appendix B.A.

Introducing more states per layer captures more accurately the spatial variability of temperature inside the layer, and thus produces a more accurate representation of the average temperature distribution as shown in Fig. 6. However, increasing the number of states in the LCM increases the computational costs and non-linearity of the LCM, which would in turn increase the training time of the PIROM. This trend is shown by the evaluation time in Fig. 6; it is two orders of magnitude more expensive to evaluate LCM-10 than LCM-3, while LCM-10 only provides a marginal improvement on the prediction accuracy. Hence LCM-10 is not considered in the rest of this study.

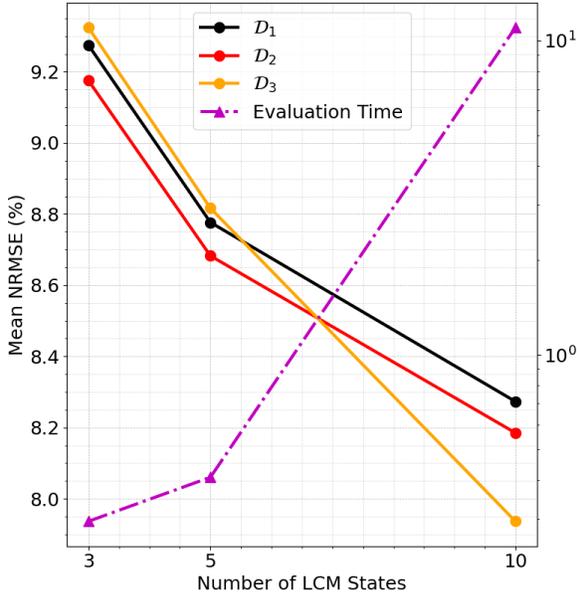
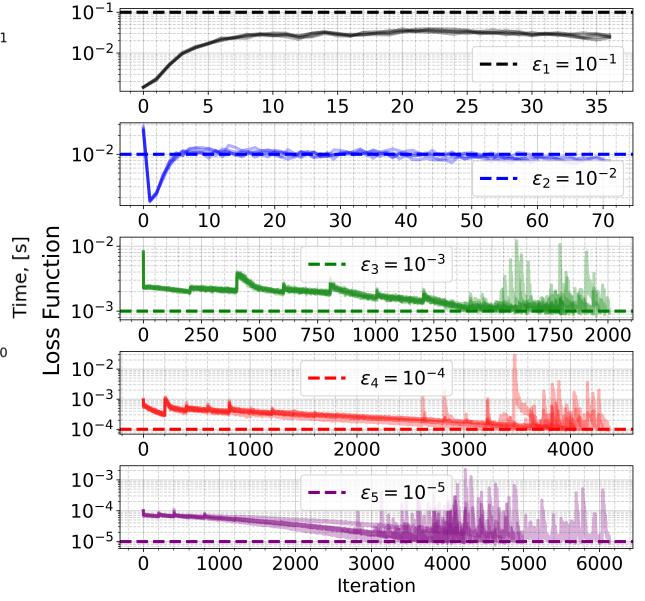

Fig. 6 Mean NRMSE and evaluation times of the LCMs.

Fig. 7 Training convergence for the PIROM with 15 total states.

### 2. Implementation and Training

The PIROM training is implemented under the `PyTorch` [40] framework using the `torchdiffeq` [6] package for the computation of the gradients in Eq. (39). Using these gradients, the `ADAM` [41] algorithm is selected to optimize the learnable parameters $\Theta$ with a learning rate of $10^{-4}$ for conservative and stable parameter search. Given the physics component and the number of hidden states, a PIROM is trained as in Algorithm 1. The hyperparameters of this algorithm are chosen as: the convergence tolerances $\mathbb{E} = \{10^{-i}\}_{i=1}^{5}$, time horizons $\mathbb{T} = \{\frac{1500}{38}j\}_{j=1}^{38}$, and maximum number of iterations $N_{\max} = 300$; the network parameters are initialized using a zero-mean isotropic Gaussian distribution with variance $10^{-4}$. All the training, except for measurement of computational costs, is conducted using 10 cores on a dual socket Intel E5-2683v3 2.00 GHz CPU with 256 GB of RAM, enabling parallel evaluation of the sensitivities for each trajectory as in Eq. (39) with negligible computational overhead.

To illustrate the effects of the tolerance-sweeping algorithm on convergence behavior, Fig. 7 presents typical training histories for the LCM-3 with 9 hidden states across five independent training runs. For the higher tolerances, $10^{-1}$ and $10^{-2}$, the training progresses rapidly to the final time horizon $t_f = 1500$ with approximately 35 and 70 iterations, respectively. Subsequently, for the lower tolerances, $10^{-3}$, $10^{-4}$, and $10^{-5}$, the optimizer becomes sensitive to the error accumulation along one trajectory, requiring substantially more iterations to achieve convergence at each time horizon stage. Notably, abrupt increases in error, such as those observed near 200 and 800 iterations in the $\epsilon = 10^{-4}$ learning curve, are attributed to the elongation of the time horizon, introducing significant trajectory mismatch that must be corrected before further progression. Without the tolerance-sweeping algorithm, direct training at the tightest tolerance level $(10^{-5})$ either incurs a prohibitive increase in training time – over two orders of magnitude higher than the current case – or fails to converge altogether.

*3. Selection of Hyperparameters*

Lastly, the hyperparameter of the PIROM is determined, which is the number of hidden states for the correction terms. A suitable number of hidden states is needed for the balance between the training time and predictive performance.

Using dataset $\mathcal{D}_1$ for training and $\mathcal{D}_2$ and $\mathcal{D}_3$ for testing, the predictive performance of the PIROM is assessed as the number of hidden states increases, as shown in Figs. 8(a) and 8(b), where the shaded area corresponds to one standard deviation of error. For a fair comparison between models based on LCM-3 and LCM-5, the total number of states in PIROM is used. The first data point, depicted as the black star, corresponds to the PIROM without hidden states, i.e., the LCM itself. As the number of hidden states increases, the mean NRMSE of the PIROMs decreases, indicating that the hidden states progressively capture more of the residual dynamics. The mean NRMSEs for the $\mathcal{D}_1$ and $\mathcal{D}_2$ overlap almost perfectly, indicating that the PIROM learns accurately the in-distribution dataset. In both cases of LCM-3 and LCM-5, when the total number of states is beyond 9, the PIROM accuracy does not significantly improve with the addition of more states; this indicates a balance between the model complexity versus the training dataset size.

Meanwhile, the computational costs of PIROM versus the number of states are shown in Fig. 9(a). For each number of states, the training is performed 5 times with different initial conditions for $\Theta_0$, and the mean and standard deviation are plotted. For the overhead ratio, shown in Fig. 9(a), the training costs decrease as the number of hidden states are increased. An explanation for this trend is that the increase in state dimension results in a better representation of the residual dynamics, and hence less iterations in training to learn the model. However, as the state dimension increases further, the computational costs increase without much improvement in predictive accuracy. For the acceleration ratio, shown in Fig. 9(b), the PIROMs based on LCM-3 are overall faster than those based on LCM-5. This difference is induced by the higher computational cost in LCM-5 when compared to LCM-3, which has been shown in Fig. 6.

Based on the above convergence study, the LCM-3 with 6 hidden states achieves a balance between predictive performance and computational efficiency. Hence this model is used exclusively in the subsequent study for ROM comparisons.

**C. Comparison of Reduced-Order Models**

In this section, all ROMs, that are trained exclusively on dataset $\mathcal{D}_1$, are assessed on varying boundary conditions and materials properties using datasets $\mathcal{D}_2$ through $\mathcal{D}_8$. Additionally, the computational efficiency of the models in terms of training and evaluation is examined.

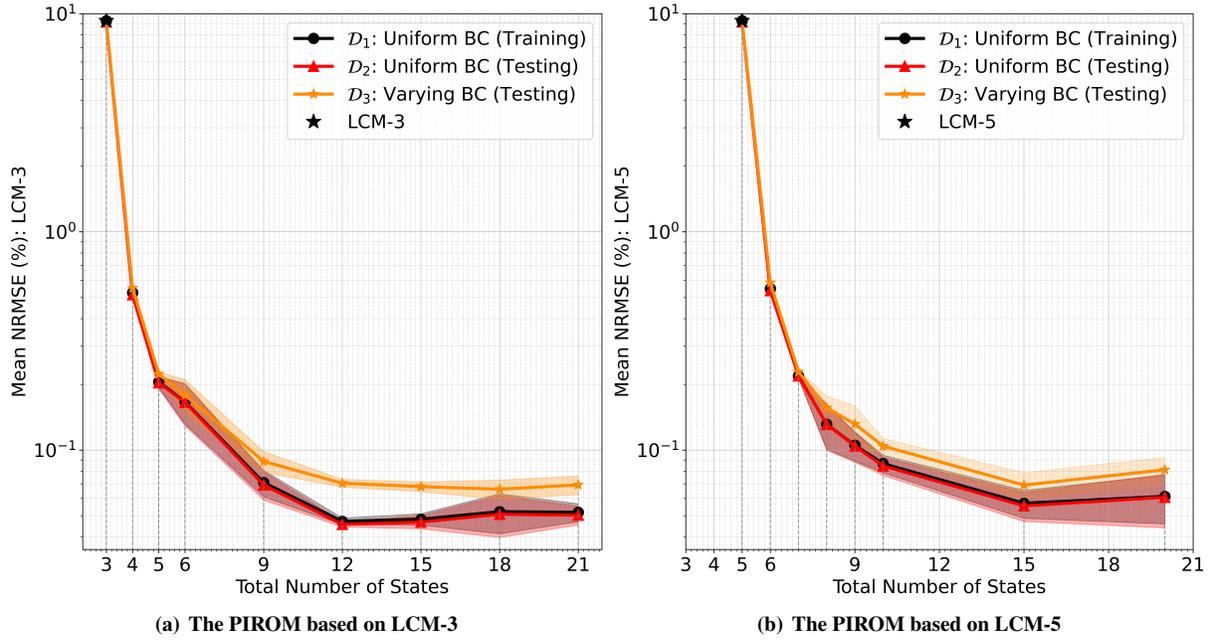

(a) **The PIROM based on LCM-3**

(b) **The PIROM based on LCM-5**

**Fig. 8  Convergence study for the number of hidden states.**

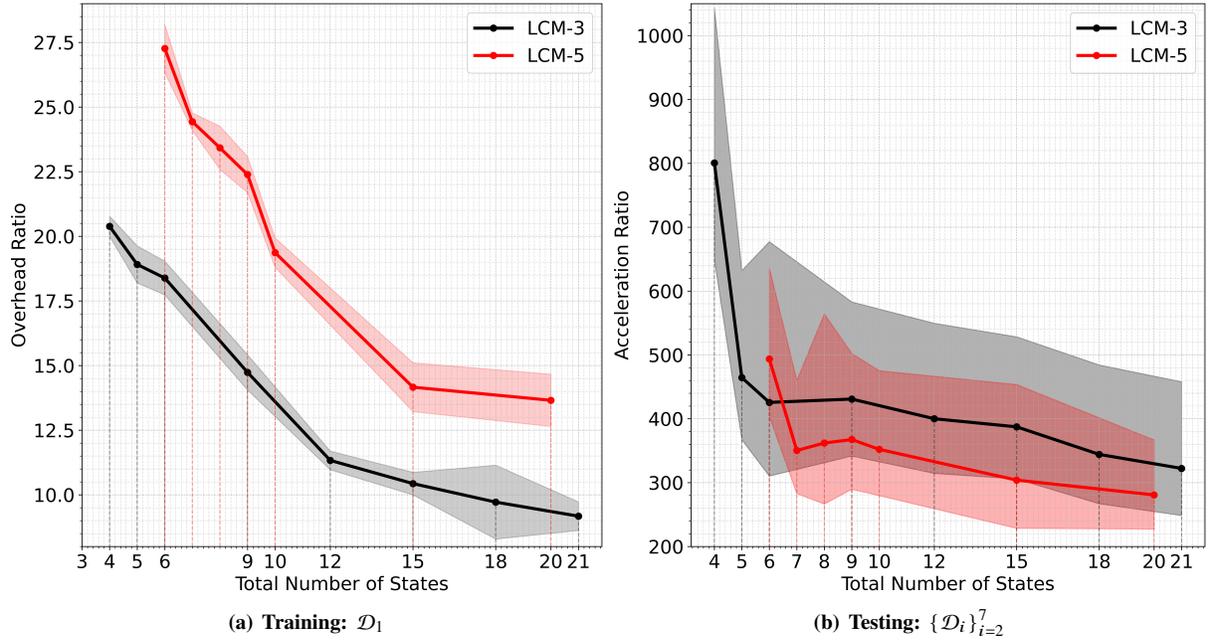

(a) **Training:** $\mathcal{D}_1$

(b) **Testing:** $\{\mathcal{D}_i\}_{i=2}^{7}$

**Fig. 9  Computational costs for the PIROMs with different states.**

*1. Generalization to Boundary Conditions*

To investigate generalization performance, the models are tested using the in-distribution dataset $\mathcal{D}_2$ and OOD datasets $\mathcal{D}_3$ to $\mathcal{D}_7$. Figure 10(a) summarizes the mean NRMSEs for each dataset; the shade shows the standard deviations of the NRMSEs. All ROMs demonstrate high prediction accuracy for the in-distribution dataset $\mathcal{D}_2$, achieving errors below 1%; this indicates appropriate training of all ROMs. As boundary condition complexity increases via parameters $\xi_1$ and $\xi_2$, prediction errors correspondingly increase for all ROMs. Notably, the OpInf and PIROM exhibit approximately one and two orders of magnitude higher accuracy compared to NODE, respectively. Such enhanced accuracy is attributed to the incorporation of physical knowledge in the ROMs.

To further characterize generalization, systematic assessments across datasets $\mathcal{D}_4 \sim \mathcal{D}_7$ are conducted with varying parameter sweeps over $\xi_1$ and $\xi_2$ for four fixed values of $\xi_0$, as illustrated in Figs. 11(a)-(d). Among the tested parameters, the NRMSE is predominantly sensitive to $\xi_2$, which controls the magnitude of the temporal Gaussian bump in the boundary conditions. The effect of $\xi_2$ is further illustrated using test trajectories in Figs. 11(e) and 11(f), denoted Case 11 and Case 95. Case 11 has a low $\xi_2$ value and resembles training data; hence all the ROMs perform relatively well. Case 95 has a high $\xi_2$ value and shows a non-linear increase-then-decrease trend in Acusil, that is not present in the training data. As a result, the NODE completely failed to capture such a non-linear trend, while OpInf and PIROM still capture the trend. This outcome is anticipated due to the purely data-driven nature of NODE, trained exclusively on cases involving uniform boundary conditions. In contrast, the physics-based structures within PIROM and OpInf effectively encode boundary condition parameters, leading to improved generalization. Specifically, Figure 11(f) demonstrates that PIROM accurately captures the temporal Gaussian boundary fluctuation, achieving an NRMSE of approximately 0.29%, whereas OpInf slightly overestimates temperature, resulting in an NRMSE of approximately 1.63%.

*2. Generalization to Material Properties*

The generalization capabilities of ROMs to material properties are evaluated using dataset $\mathcal{D}_8$. Given that OpInf and NODE models are formulated for boundary condition generalization, they are inherently unable to adapt to variations in material properties without further parametrization. Consequently, only PIROM is examined in this context.

Figure 12 illustrates a representative case of PIROM generalizing to material property variations, where Fig. 12(a) illustrates the nominal and perturbed material properties and Fig. 12(b) compares the full-order and PIROM predictions in both nominal and perturbed cases. Note again that PIROM is trained using only the nominal material properties. Although the boundary conditions remain identical across these scenarios, perturbations in material properties elevate the temperature loads within the layers. Nonetheless, PIROM effectively incorporates the effects of these perturbations through adjustments in the LCM mass and stiffness matrices, enabling accurate observable predictions with minimal errors.

The overall mean NRMSE values for PIROM across all cases in dataset $\mathcal{D}_8$ are presented in Figure 10, along with one standard deviation. Results indicate that PIROM consistently achieves a low NRMSE of approximately 1%, demonstrating robust generalization capability even without specific training data or additional parametrization for the perturbed materials. This performance is attributed to the physics-infusion formulation, which effectively captures and compensates for the non-linear material-induced dynamics.

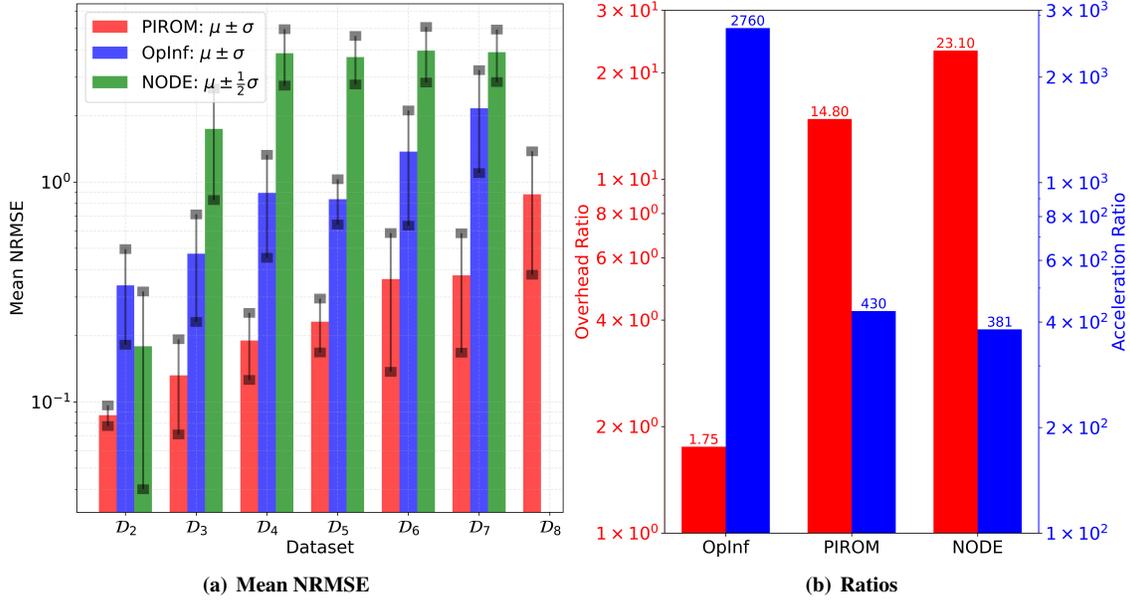

Fig. 10   Prediction and training performance of the ROMs.

*3. Computational Cost*

The computational costs of the ROMs are compared in Fig. 10(b), where the acceleration ratio is evaluated using datasets $\mathcal{D}_1$ and the overhead ratio is evaluated using datasets $\mathcal{D}_2 \sim \mathcal{D}_7$. Note that in the comparison serial computation is assumed for fairness.

The OpInf achieves the highest acceleration ratio, outperforming both PIROM and NODE by approximately one order of magnitude. This difference stems from the underlying numerical solvers: OpInf integrates reduced-order dynamics using `SciPy`'s compiled solvers (e.g., LSODA), which draw upon highly optimized Fortran/C libraries. In contrast, both PIROM and NODE rely on `torchdiffeq`'s differentiable solvers, which are natively written using the `PyTorch` library and thus prioritize flexibility and backpropagation compatibility over computational efficiency.

Regarding training cost, PIROM and NODE exhibit significantly higher overhead ratios than OpInf. This is primarily due to the need to solve adjoint-based trajectory optimization problems, as formulated in Eq. (36). For PIROM, this translates into substantial computational expense to achieve the high generalizability and accuracy demonstrated in preceding sections. However, this cost is amenable to several mitigation strategies. First, gradient evaluations based on adjoint sensitivities may be massively parallelized across trajectories. Second, training efficiency may be improved

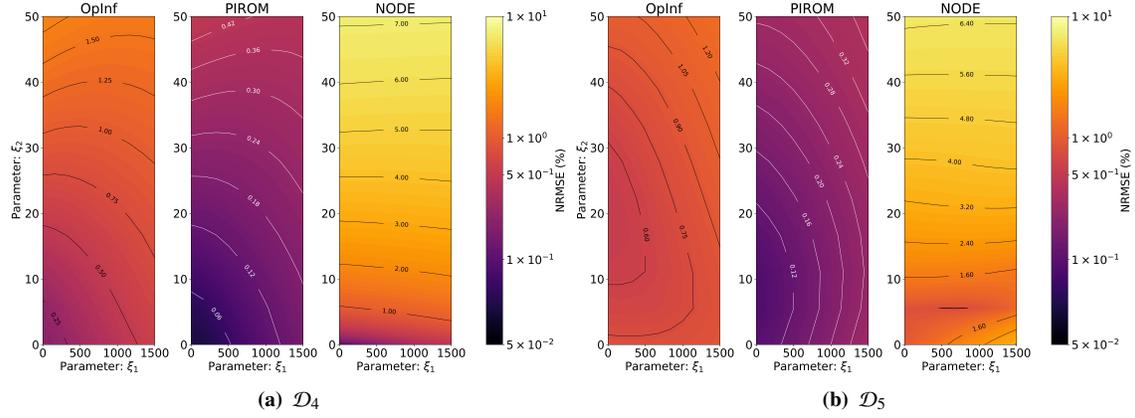

(a) $\mathcal{D}_4$      (b) $\mathcal{D}_5$

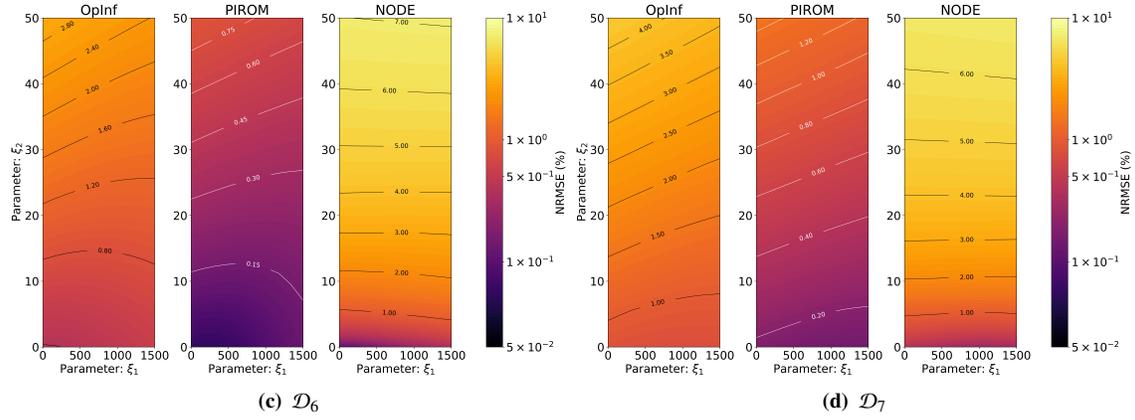

(c) $\mathcal{D}_6$      (d) $\mathcal{D}_7$

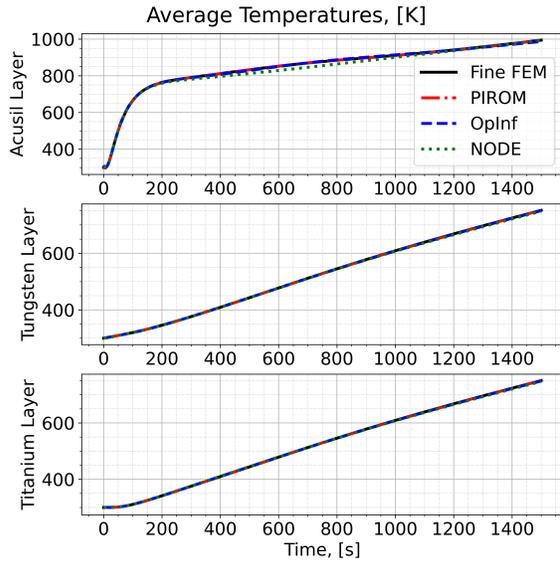
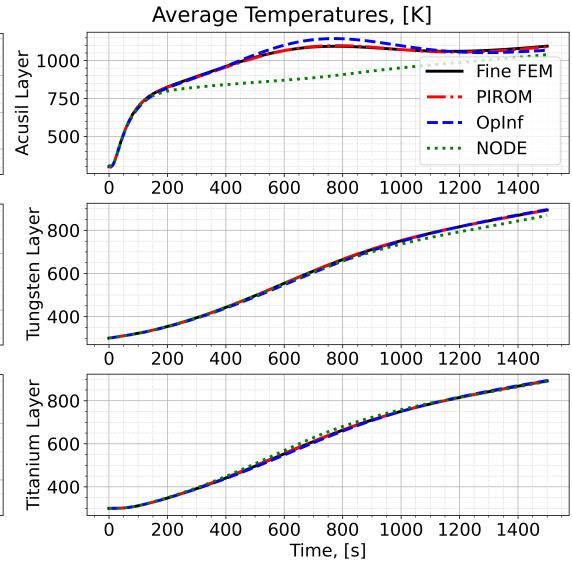

(e) **Test case 11 in $\mathcal{D}_4$ with low $\xi_2$**      (f) **Test Case 95 in $\mathcal{D}_4$ with high $\xi_2$**

**Fig. 11**   **The NRMSE distribution for the BC sweeps.**

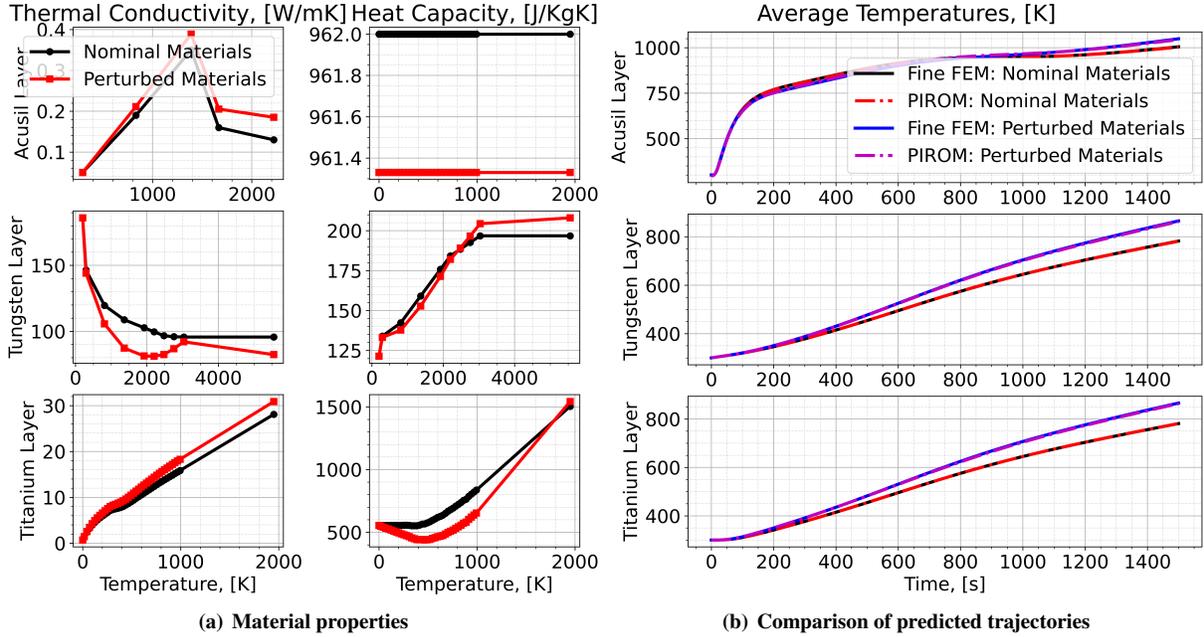

(a) **Material properties**

(b) **Comparison of predicted trajectories**

Fig. 12    Performance of PIROM under new material properties.

through targeted hyperparameter tuning; for example, the $10^{-1}$ and $10^{-2}$ tolerances used in Sec. IV.B.2 may be removed as the optimizer is insensitive to them. Third, recent advances in weak-form training methods, such as those in Ref. [42], offer promising acceleration strategies. In particular, Ref. [42] demonstrates that adjoint-free weak-form training algorithms can reduce training time for neural ODEs by up to three orders of magnitude.

*4. Summary of Comparison*

The study above highlights three key findings: (1) incorporating physics into data-driven modeling significantly enhances generalizability without increasing data requirements; (2) the physics-based component in PIROM enables robust handling of problem parametrization; and (3) PIROM balances superior generalizability and accuracy at the expense of increased training cost.

First, the PIROM's improved generalization is primarily due to the low-order LCM, which provides a physics-consistent baseline prediction. The PIROM then augments this prediction by introducing hidden data-driven states to capture higher-order residual dynamics. Using a different approach, the OpInf incorporates physics via projection, allowing some generalization to new boundary conditions, but its fixed operators lack parametric dependence and require additional modeling effort to handle material variability. The NODE, being purely data-driven, fails to generalize outside its training regime, demonstrating the limitations of physics-agnostic models in extrapolative tasks.

Second, PIROM's architecture is designed to partition the modeling burden: the LCM enforces parametrically-compatible fundamental physical laws, while the hidden states selectively correct its deficiencies. This division ensures that the model is both physically interpretable and computationally efficient, avoiding the redundancy of re-learning known physics from data.

Third, while PIROM incurs in higher computational cost during training compared to OpInf, this expense is a one-time investment. Moreover, the training overhead is amenable to significant reduction through parallelized optimization strategies and algorithmic advances such as weak-form learning. Consequently, PIROM offers a scalable and high-fidelity modeling framework that effectively balances generalization, accuracy, and efficiency.

## V. Conclusion

This study presents a physics-infused reduced-order modeling (PIROM) framework and applies it to model the transient thermal response of multi-layered hypersonic thermal protection systems (TPS). The first objective – formulating the PIROM – is achieved by deriving a coarse-grained model grounded in the Mori-Zwanzig formalism. Specifically, it was shown that the coarse-graining of a full-order model (DG FEM) exactly produces a reduced-physics model (LCM), together with a term of residual dynamics. Using Mori-Zwanzig formalism, the latter is replaced by a linear system of hidden states. The result is a reduced-order structure composed of a low-order LCM that captures the leading-order thermal dynamics, and a data-driven correction term that resolves residual, higher-order behaviors. The resulting model retains physical interpretability while enabling generalization across a wide range of boundary conditions and material property variations.

To address the second objective, a comprehensive convergence analysis is conducted. Results demonstrate that as the number of hidden states increases, the PIROM solutions convergence systematically toward the FOM, achieving less than 1% NRMSE across all test cases. This validates the theoretical underpinnings of the PIROM framework and confirms its ability to approximate high-fidelity models with controllable accuracy. Importantly, the PIROM requires only thermocouple temperature histories for training, in contrast to methods such as OpInf, which rely on full-field temperature snapshots. Such distinction significantly reduces the data acquisition burden and enhances the practicality of PIROM for experimental or in-flight conditions.

The third objective evaluates the PIROM relative to OpInf and NODE along the three axes of the impossible trinity of modeling: accuracy, generalizability, and efficiency. The results show that PIROM clearly resolves the accuracy and generalizability challenges: it consistently achieves the lowest extrapolation errors to boundary conditions, including non-linear material perturbations unseen during training. This performance stems from its physics-infused backbone and its data-driven correction mechanism. As for the computational efficiency, PIROM does exhibit accelerations of two orders of magnitude in prediction, and its forward solve times are sufficiently low for many-query tasks. Indeed, PIROM is slower to evaluate than OpInf, but this is primarily due to software implementation differences, and the latter still suffers from the generalizability. In short, the PIROM nearly resolves the efficiency axis during evaluation.

The principal limitation of PIROM remains in the training costs, where the PIROM incurs a higher computational burden due to its reliance on trajectory optimization with continuous adjoint-based gradient evaluations, including massively parallelization and the weak-form learning [42]. With these improvements, PIROM has the potential to become a highly competitive framework that approaches all three vertices of the impossible trinity – combining accuracy, generalizability, and efficiency in a single, scalable modeling architecture.

Future research efforts along the lines of TPS will extend PIROM along both application and algorithmic dimensions. On the application side, the PIROM can be generalized to handle ablation-induced surface recession, introducing additional governing equations and coupling challenges due to evolving material boundaries; the PIROM can also be applied for the uncertainty quantification and design optimization of the TPS. On the algorithmic side, the aforementioned learning acceleration techniques will be prioritized to reduce training overhead, and the theoretical convergence bounds of PIROM also worth further investigation. Together, these developments position PIROM as a strong candidate for resolving the long-standing impossible trinity of modeling in high-fidelity simulation of extreme aerospace environments.

## VI. Acknowledgments

The authors would like to acknowledge Shane McQuarrie for his help on setting up the Operator Inference ROM. D.H. acknowledges the support from the NSF CAREER program CMMI-2340266.

Sandia National Laboratories is a multi-mission laboratory managed and operated by National Technology & Engineering Solutions of Sandia, LLC (NTESS), a wholly owned subsidiary of Honeywell International Inc., for the U.S. Department of Energy's National Nuclear Security Administration (DOE/NNSA) under contract DE-NA0003525. This written work is authored by an employee of NTESS. The employee, not NTESS, owns the right, title and interest in and to the written work and is responsible for its contents. Any subjective views or opinions that might be expressed in the written work do not necessarily represent the views of the U.S. Government. The publisher acknowledges that the U.S. Government retains a non-exclusive, paid-up, irrevocable, world-wide license to publish or reproduce the published form of this written work or allow others to do so, for U.S. Government purposes. The DOE will provide public access to results of federally sponsored research in accordance with the DOE Public Access Plan.

## A. The Coarse-Graining Procedure

This appendix provides the details of coarse-graining for the TPS problem, the magnitude analysis of residual dynamics after the coarse-graining, and an illustrative 1D example.

## A. Detailed Derivation of Coarse-Graining

This section supplements the details pertaining to Eqs. (22) and (23). First, by definition,

$$
\begin{aligned}
\mathbf{r}^{(1)}(\bar{\mathbf{u}}, t) &= \mathcal{P}\left[\mathbf{\Phi}^+\mathbf{r}(\mathbf{u}, t)\right] \\
&= \mathcal{P}\left[\mathbf{\Phi}^+\mathbf{A}(\mathbf{u})^{-1}\mathbf{B}(\mathbf{u})\mathbf{u} + \mathbf{\Phi}^+\mathbf{A}(\mathbf{u})^{-1}\mathbf{f}(t)\right] \\
&= \mathbf{\Phi}^+\mathbf{A}(\mathbf{P}\mathbf{u})^{-1}\mathbf{P}\mathbf{B}(\mathbf{P}\mathbf{u})\mathbf{P}\mathbf{u} + \mathbf{\Phi}^+\mathbf{A}(\mathbf{P}\mathbf{u})^{-1}\mathbf{P}\mathbf{f}(t) \\
&= \underbrace{\left(\mathbf{\Phi}^+\mathbf{A}(\mathbf{\Phi}\bar{\mathbf{u}})^{-1}\mathbf{\Phi}\right)}_{\#1}\underbrace{\left(\mathbf{\Phi}^+\mathbf{B}(\mathbf{\Phi}\bar{\mathbf{u}})\mathbf{\Phi}\right)}_{\#2}\bar{\mathbf{u}} + \left(\mathbf{\Phi}^+\mathbf{A}(\mathbf{\Phi}\bar{\mathbf{u}})^{-1}\mathbf{\Phi}\right)\underbrace{\left(\mathbf{\Phi}^+\mathbf{f}(t)\right)}_{\#3},
\end{aligned}
\quad (44)
$$

where the terms #1, #2, and #3 are the coarse-graining of the (inverted) heat capacitance matrix, the heat conduction matrix, and the heat load, respectively. Next, these terms are treated one by one.

**Term #1** The $\mathbf{A}$ matrix is block-diagonal, and as discussed in Eq. (8), a diagonal block $\mathbf{A}_i$ itself is diagonal if the material property $\rho c_p$ is constant within the $i$th element. Specifically, take an element $E_i \in \Omega_j$ and denote $\overline{\rho c}_{p,j}$ as the value of $\rho c_p$ at $\bar{u}_j$, then $\mathbf{A}_i = \overline{\rho c}_{p,j}|E_i|\mathbf{I}$, where $|E_i|$ is the size of the $i$th element.

Then, the $(k, l)$th element of term #1 is computed as

$$
\begin{aligned}
\left[\mathbf{\Phi}^+\mathbf{A}(\mathbf{\Phi}\bar{\mathbf{u}})^{-1}\mathbf{\Phi}\right]_{kl} &= \sum_{i=1}^{M}\sum_{j=1}^{M} \boldsymbol{\varphi}_i^{k+}[\mathbf{A}(\mathbf{\Phi}\bar{\mathbf{u}})^{-1}]_{ij}\boldsymbol{\varphi}_j^l \\
&= \sum_{i=1}^{M}\sum_{j=1}^{M} \boldsymbol{\varphi}_i^{k+}\left[(\overline{\rho c}_{p,k}|E_i|)^{-1}\mathbf{I}\delta_{ij}\right]\boldsymbol{\varphi}_j^l \\
&= \sum_{i=1}^{M} \boldsymbol{\varphi}_i^{k+}(\overline{\rho c}_{p,k}|E_i|)^{-1}\boldsymbol{\varphi}_i^l \\
&= \sum_{i \in \mathcal{V}_k} \boldsymbol{\varphi}_i^{k+}(\overline{\rho c}_{p,k}|E_i|)^{-1}\boldsymbol{\varphi}_i^l,
\end{aligned}
\quad (45)
$$

where in the second row $\delta_{ij}$ is Kronecker delta, in the third row the summation property of $\delta_{ij}$ is applied, in the fourth row the summation is reduced to $i \in \mathcal{V}_k$ since $\boldsymbol{\varphi}_i^k = 0$ if $i \neq \mathcal{V}_k$.

Subsequently, if $k \neq l$, then $i \notin \mathcal{V}_l$, $\boldsymbol{\varphi}_i^l = 0$, and the term in Eq. (45) is zero. When $k = l$, by definition

$$
\left[\mathbf{\Phi}^+\mathbf{A}(\mathbf{\Phi}\bar{\mathbf{u}})^{-1}\mathbf{\Phi}\right]_{kl} = \sum_{i \in \mathcal{V}_k} \frac{|E_i|}{|\Omega_k|}\boldsymbol{\varphi}_i^{k\top}(\overline{\rho c}_{p,k}|E_i|)^{-1}\boldsymbol{\varphi}_i^k = \sum_{i \in \mathcal{V}_k}(\overline{\rho c}_{p,k}|\Omega_k|)^{-1} = |\mathcal{V}_k|(\overline{\rho c}_{p,k}|\Omega_k|)^{-1}.
$$

Hence, if one introduces a diagonal weight matrix $\mathbf{W} \in \mathbb{R}^{M \times M}$, such that the $i$th diagonal element $[\mathbf{W}]_i = |\mathcal{V}_k|$ if $i \in \mathcal{V}_k$, then clearly

$$
\mathbf{\Phi}^+\mathbf{A}(\mathbf{\Phi}\bar{\mathbf{u}})^{-1}\mathbf{\Phi}\mathbf{W}^{-1} = \bar{\mathbf{A}}(\bar{\mathbf{u}})^{-1}, \text{ or } \mathbf{W}\left(\mathbf{\Phi}^+\mathbf{A}(\mathbf{\Phi}\bar{\mathbf{u}})^{-1}\mathbf{\Phi}\right)^{-1} = \bar{\mathbf{A}}(\bar{\mathbf{u}}),
$$

as stated in Eq. (23).

**Term #2**  The $\mathbf{B}$ matrix has a block structure, where the $(i,j)$th block $\mathbf{B}_{ij} \neq 0$ if $j \in \mathcal{N}_i$. Since the first basis function was chosen to be a constant, the $(1,1)$th element of $\mathbf{B}_{ij}$ is simply

$$[\mathbf{B}_{ij}]_{11} = \begin{cases} -\sum_{l \in \mathcal{N}_i \cup \{T\}} \sigma |e_{il}|, & i = j, \\ \sigma |e_{ij}|, & j \in \mathcal{N}_i, \\ 0, & \text{otherwise}. \end{cases} \quad (46)$$

Then, the $(k,l)$th element of term #2 is computed as

$$\left[\mathbf{\Phi}^+ \mathbf{B}(\mathbf{\Phi}\bar{\mathbf{u}}) \mathbf{\Phi}\right]_{kl} = \sum_{i=1}^{M} \sum_{j=1}^{M} \varphi_i^{k+} \mathbf{B}_{ij} \varphi_j^l = \sum_{i \in \mathcal{V}_k} \sum_{j \in \mathcal{V}_l} \varphi_i^{k+} \mathbf{B}_{ij} \varphi_j^l$$

$$= \sum_{i \in \mathcal{V}_k} \sum_{j \in \mathcal{V}_l} \frac{|E_i|}{|\Omega_k|} \varphi_i^{k\top} \mathbf{B}_{ij} \varphi_j^l = \frac{1}{|\mathcal{V}_k|} \sum_{i \in \mathcal{V}_k} \sum_{j \in \mathcal{V}_l} [\mathbf{B}_{ij}]_{11}, \quad (47)$$

where the last step introduced a mild assumption that the element sizes are the same within a component $\Omega_k$.

To simplify Eq. (47) further, note the special $+/-$ structure in Eq. (46); this implies that if an edge $e_{ij}$ is not on the boundary of a component, then the $\sigma |e_{ij}|$ terms will cancel upon summation. Eventually, only the edges that are adjacent to another component or belong to the BC will remain after summation in Eq. (47).

To perform the subsequent calculations, the following notations are introduced: (1) The collection of edges between components $k$ and $l$: $\mathcal{E}_{kl} = \{(i,j) | j \in \mathcal{N}_i, i \in \mathcal{V}_k, j \in \mathcal{V}_l\}$; (2) The collection of edges between components $k$ and Dirichlet BC: $\mathcal{E}_{kT} = \{(i,T) | i \in \mathcal{V}_k\}$; (3) The lengths of boundaries: $|e_{kl}| = \sum_{(i,j) \in \mathcal{E}_{kl}} |e_{ij}|$, $|e_{kT}| = \sum_{(i,T) \in \mathcal{E}_{kT}} |e_{iT}|$; (4) The neighbors of component $k$: $\bar{\mathcal{N}}_k = \{l | \mathcal{E}_{kl} \neq \emptyset\}$.

The summation in Eq. (47) is simplified as

$$\sum_{i \in \mathcal{V}_k} \sum_{j \in \mathcal{V}_l} [\mathbf{B}_{ij}]_{11} = \begin{cases} -\sum_{p \in \bar{\mathcal{N}}_k \cup \{T\}} \sum_{(i,j) \in \mathcal{E}_{kp}} \sigma |e_{ij}| = -\sum_{p \in \bar{\mathcal{N}}_k \cup \{T\}} \sigma |e_{kp}|, & k = l, \\ \sum_{(i,j) \in \mathcal{E}_{kl}} \sigma |e_{ij}| = \sigma |e_{kl}|, & l \in \bar{\mathcal{N}}_k, \\ 0, & \text{otherwise} \end{cases} \quad (48)$$

which are exactly the $\bar{\mathbf{B}}$ matrix in LCM. In other words, one obtains

$$\mathbf{W}\mathbf{\Phi}^+ \mathbf{B}(\mathbf{\Phi}\bar{\mathbf{u}}) \mathbf{\Phi} = \bar{\mathbf{B}}(\bar{\mathbf{u}}),$$

as stated in Eq. (23).

**Term #3** Consider an element $i$ with an edge $e_{iq}$ on the Neumann BC. Due to the first basis function as a constant, the first entry of $\mathbf{f}_i$ is

$$[\mathbf{f}_i]_1 = |e_{iq}|\bar{q}_{b,i},$$

where $\bar{q}_{b,i}$ is the average heat flux over $e_{iq}$.

Then, the $k$th entry of term #3 is computed as

$$[\boldsymbol{\Phi}^+\mathbf{f}]_k = \sum_{i\in\mathcal{V}_k} \boldsymbol{\varphi}_i^{k+}\mathbf{f}_i = \sum_{i\in\mathcal{V}_k} \frac{|E_i|}{|\Omega_k|}\boldsymbol{\varphi}_i^{k\top}\mathbf{f}_i = \sum_{i\in\mathcal{V}_k} \frac{|E_i|}{|\Omega_k|}[\mathbf{f}_i]_1$$
$$= \frac{1}{|\mathcal{V}_k|}\sum_{i\in\mathcal{V}_k} |e_{iq}|\bar{q}_{b,i} = \frac{1}{|\mathcal{V}_j|}|e_{kq}|\bar{q}_{b,k},$$

where the mild assumption of equal element size within $\mathcal{V}_j$ is invoked again, and in the last step $|e_{kq}|$ is the length of boundary of component $k$ on the Neumann BC, and $\bar{q}_{b,k}$ is the average heat flux over $e_{kq}$.

Comparing with the LCM, one obtains

$$\mathbf{W}\boldsymbol{\Phi}^+\mathbf{f} = \bar{\mathbf{f}},$$

as stated in Eq. (23).

**Summary** Back to Eq. (44), the resolved dynamics, Eq. (22), is recovered as follows,

$$\mathbf{r}^{(1)}(\bar{\mathbf{u}}, t) = \left(\boldsymbol{\Phi}^+\mathbf{A}(\boldsymbol{\Phi}\bar{\mathbf{u}})^{-1}\boldsymbol{\Phi}\right)\left(\boldsymbol{\Phi}^+\mathbf{B}(\boldsymbol{\Phi}\bar{\mathbf{u}})\boldsymbol{\Phi}\right)\bar{\mathbf{u}} + \left(\boldsymbol{\Phi}^+\mathbf{A}(\boldsymbol{\Phi}\bar{\mathbf{u}})^{-1}\boldsymbol{\Phi}\right)\left(\boldsymbol{\Phi}^+\mathbf{f}(t)\right)$$
$$= \left(\boldsymbol{\Phi}^+\mathbf{A}(\boldsymbol{\Phi}\bar{\mathbf{u}})^{-1}\boldsymbol{\Phi}\right)\mathbf{W}^{-1}\mathbf{W}\left(\boldsymbol{\Phi}^+\mathbf{B}(\boldsymbol{\Phi}\bar{\mathbf{u}})\boldsymbol{\Phi}\right)\bar{\mathbf{u}} + \left(\boldsymbol{\Phi}^+\mathbf{A}(\boldsymbol{\Phi}\bar{\mathbf{u}})^{-1}\boldsymbol{\Phi}\right)\mathbf{W}^{-1}\mathbf{W}\left(\boldsymbol{\Phi}^+\mathbf{f}(t)\right)$$
$$= \bar{\mathbf{A}}(\bar{\mathbf{u}})^{-1}\bar{\mathbf{B}}(\bar{\mathbf{u}})\bar{\mathbf{u}} + \bar{\mathbf{A}}(\bar{\mathbf{u}})^{-1}\bar{\mathbf{f}}(t).$$

## B. Magnitude Analysis of Residual Dynamics

Next, since the residual dynamics $\mathbf{r}^{(2)}$ encompasses the physics that are missing in the resolved dynamics $\mathbf{r}^{(1)}$, i.e., the LCM, the terms in $\mathbf{r}^{(2)}$ are examined in more detail to pinpoint these missing physics.

By definition

$$\mathbf{r}^{(2)}(\mathbf{u}, t) = \mathbf{r}(\mathbf{u}, t) - \mathbf{r}^{(1)}(\bar{\mathbf{u}}, t) = \left(\boldsymbol{\Phi}^+\mathbf{A}^{-1}\mathbf{B}\mathbf{u} - \bar{\mathbf{A}}^{-1}\bar{\mathbf{B}}\bar{\mathbf{u}}\right) + \left(\boldsymbol{\Phi}^+\mathbf{A}^{-1}\mathbf{f} - \bar{\mathbf{A}}^{-1}\bar{\mathbf{f}}\right).$$

For the first group of terms,

$$\|\mathbf{\Phi}^+\mathbf{A}^{-1}\mathbf{B}\mathbf{u} - \bar{\mathbf{A}}^{-1}\bar{\mathbf{B}}\bar{\mathbf{u}}\| = \|\mathbf{\Phi}^+\mathbf{A}(\mathbf{u})^{-1}\mathbf{B}(\mathbf{u})\mathbf{u} - \mathbf{\Phi}^+\mathbf{A}(\mathbf{\Phi}\bar{\mathbf{u}})^{-1}\mathbf{P}\mathbf{B}(\mathbf{\Phi}\bar{\mathbf{u}})\mathbf{\Phi}\mathbf{\Phi}^+\mathbf{u}\|$$

$$\leq \|\mathbf{\Phi}^+\mathbf{A}(\mathbf{u})^{-1}\mathbf{B}(\mathbf{u})\mathbf{u} - \mathbf{\Phi}^+\mathbf{M}(\bar{\mathbf{u}})\mathbf{u}\| + \|\mathbf{\Phi}^+\mathbf{M}(\bar{\mathbf{u}})\mathbf{u} - \mathbf{\Phi}^+\mathbf{M}(\bar{\mathbf{u}})\mathbf{\Phi}\bar{\mathbf{u}}\|$$

$$\leq \|\mathbf{\Phi}^+\|\underbrace{\|\mathbf{A}(\mathbf{u})^{-1}\mathbf{B}(\mathbf{u}) - \mathbf{M}(\bar{\mathbf{u}})\|}_{\#1}\|\mathbf{u}\| + \|\mathbf{\Phi}^+\mathbf{M}(\bar{\mathbf{u}})\|\underbrace{\|\mathbf{u} - \mathbf{\Phi}\bar{\mathbf{u}}\|}_{\#2},$$

where the second row introduces a simplifying notation $\mathbf{M}(\bar{\mathbf{u}}) = \mathbf{A}(\mathbf{\Phi}\bar{\mathbf{u}})^{-1}\mathbf{P}\mathbf{B}(\mathbf{\Phi}\bar{\mathbf{u}})$ and used triangular inequality, and the third row factors out common terms. The differences are labeled #1 and #2. Term #2 is directly due to the approximation of non-uniform temperature as constants, and term #1 is the error in material properties due to such an approximation.

For the second group of terms, a similar procedure is applied,

$$\|\mathbf{\Phi}^+\mathbf{A}^{-1}\mathbf{f} - \bar{\mathbf{A}}^{-1}\bar{\mathbf{f}}\| = \|\mathbf{\Phi}^+\mathbf{A}(\mathbf{u})^{-1}\mathbf{f} - \mathbf{\Phi}^+\mathbf{A}(\mathbf{\Phi}\bar{\mathbf{u}})^{-1}\mathbf{\Phi}\mathbf{\Phi}^+\mathbf{f}\|$$

$$\leq \|\mathbf{\Phi}^+\mathbf{A}(\mathbf{u})^{-1}\mathbf{f} - \mathbf{\Phi}^+\mathbf{A}(\mathbf{\Phi}\bar{\mathbf{u}})^{-1}\mathbf{f}\| + \|\mathbf{\Phi}^+\mathbf{A}(\mathbf{\Phi}\bar{\mathbf{u}})^{-1}\mathbf{f} - \mathbf{\Phi}^+\mathbf{A}(\mathbf{\Phi}\bar{\mathbf{u}})^{-1}\mathbf{\Phi}\bar{\mathbf{f}}\|$$

$$\leq \|\mathbf{\Phi}^+\|\underbrace{\|\mathbf{A}(\mathbf{u})^{-1} - \mathbf{A}(\mathbf{\Phi}\bar{\mathbf{u}})^{-1}\|}_{\#1}\|\mathbf{f}\| + \|\mathbf{\Phi}^+\mathbf{A}(\mathbf{\Phi}\bar{\mathbf{u}})^{-1}\|\underbrace{\|\mathbf{f} - \mathbf{\Phi}\bar{\mathbf{f}}\|}_{\#2},$$

where term #1 is again the error in material properties due to the piece-wise constant approximation of non-uniform temperature, and term #2 is due to the piece-wise constant approximation of the heat load.

## C. A One-Dimensional Example of Coarse-Graining

Consider the one-dimensional domain with two non-overlapping components $N = 2$, where $\Omega = \Omega_1 \cup \Omega_2$ and $\Omega_i = [x_i, x_{i+1}]$ for $i \in \{1, 2\}$ as shown in Fig. 13. The governing conduction equation is,

$$\rho c_p \frac{\partial T}{\partial t} = \frac{\partial}{\partial x}\left(k(x)\frac{\partial T}{\partial x}\right) \tag{49a}$$

$$-k\frac{\partial T}{\partial x}\bigg|_{x_1} = q_1(t) \tag{49b}$$

$$-k\frac{\partial T}{\partial x}\bigg|_{x_3} = q_3(t) \tag{49c}$$

$$T(x, 0) = T_0(x), \quad x \in \Omega. \tag{49d}$$

where the material properties are piecewise-continuous functions of $x$, with possible discontinuity at the component interface $x = x_2$.

For DG, a single first-order finite element is assigned to each component, i.e., $E_1 = \Omega_1$ and $E_2 = \Omega_2$, and thus $N = 2$

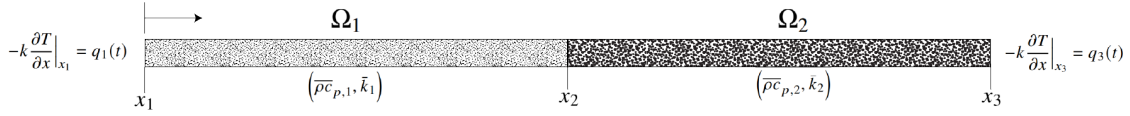

Fig. 13  One-dimensional composite domain with two finite elements.

and $P = 2$. Over the $i$-th element, the states $\mathbf{u}_i \in \mathbb{R}^2$ are associated with the following orthogonal polynomial basis,

$$\phi_1^i(x) = \begin{cases} 1, & x \in E_i \\ 0, & \text{otherwise} \end{cases}, \quad \phi_2^i(x) = \begin{cases} \sqrt{3}\left(1 - \frac{2}{|E_i|}(x - x_i)\right), & x \in E_i \\ 0, & \text{otherwise} \end{cases} \quad (50)$$

where $|E_i| = x_{i+1} - x_i$.

For simplicity, assume the material properties $\rho c_p$ and $k$ are constants evaluated at the average temperature $\bar{u}$, i.e., $\rho c_p(\mathbf{u}) = \rho c_p(\bar{u}) = \overline{\rho c}_p$ and $k(\mathbf{u}) = k(\bar{u}) = \bar{k}$. Thus, using Eq. (8) the DG operators become,

$$\mathbf{A}(\mathbf{u}) = \begin{pmatrix} \alpha_1 & 0 & 0 & 0 \\ 0 & \alpha_1 & 0 & 0 \\ 0 & 0 & \alpha_2 & 0 \\ 0 & 0 & 0 & \alpha_2 \end{pmatrix}, \quad (51a)$$

$$\mathbf{B}(\mathbf{u}) = \begin{pmatrix} -\sigma & \sigma & \sigma & \sigma \\ \sqrt{3}\gamma_1 & (6-\sqrt{3})\gamma_1 - \sigma & -\sqrt{3}\gamma_1 - \sigma & -\sqrt{3}\gamma_1 - \sigma \\ \sigma & \sigma & -\sigma & -\sigma \\ -\sqrt{3}\gamma_2 - \sigma & \sqrt{3}\gamma_2 + \sigma & \sqrt{3}\gamma_2 - \sigma & (6+\sqrt{3})\gamma_2 - \sigma \end{pmatrix}, \quad \mathbf{f}(t) = \begin{pmatrix} q_1(t) \\ q_1(t) \\ q_3(t) \\ -q_3(t) \end{pmatrix}. \quad (51b)$$

where $\alpha_i = \overline{\rho c}_{p,i}|E_i|$ and $\gamma_i = 2\bar{k}_i/|E_i|$.

Next, the states are coarse-grained as in Eq. (19), with

$$\mathbf{\Phi}^+ = \begin{pmatrix} \frac{|E_1|}{|\Omega_1|} & 0 & 0 & 0 \\ 0 & 0 & \frac{|E_2|}{|\Omega_2|} & 0 \end{pmatrix}, \quad \mathbf{\Phi} = \begin{pmatrix} 1 & 0 \\ 0 & 0 \\ 0 & 1 \\ 0 & 0 \end{pmatrix}, \quad (52)$$

where each component has only one element, so $|E_i|/|\Omega_i| = 1$. Performing the projections for the term #1 in the previous section yields,

$$\bar{\mathbf{A}} = \begin{pmatrix} |\mathcal{V}_1||\overline{\rho c}_{p,1}|\Omega_1| & 0 \\ 0 & |\mathcal{V}_2||\overline{\rho c}_{p,2}|\Omega_2| \end{pmatrix} = \begin{pmatrix} \overline{\rho c}_{p,1}|\Omega_1| & 0 \\ 0 & \overline{\rho c}_{p,2}|\Omega_2| \end{pmatrix}, \quad (53)$$

where $|\mathcal{V}_i| = 1$ for $i = 1, 2$. The formula for term #2 yields,

$$\bar{\mathbf{B}}(\bar{\mathbf{u}}) = \begin{pmatrix} -\sigma|e_{11}| & \sigma|e_{12}| \\ \sigma|e_{21}| & -\sigma|e_{22}| \end{pmatrix} = \begin{pmatrix} -\frac{|e_{11}|}{R_{12}} & \frac{|e_{12}|}{R_{12}} \\ \frac{|e_{21}|}{R_{12}} & -\frac{|e_{22}|}{R_{12}} \end{pmatrix}, \quad (54)$$

where the penalty factor $\sigma = \frac{1}{R_{12}}$ is set to be the inverse of the equivalent thermal resistance between the two neighboring materials.

Lastly the formula for the term #3 yields,

$$\bar{\mathbf{f}} = \begin{pmatrix} |\mathcal{V}_1|\frac{|E_1|}{|\Omega_1|}|e_{1q}|\bar{q}_1(t) \\ |\mathcal{V}_2|\frac{|E_2|}{|\Omega_2|}|e_{2q}|\bar{q}_3(t) \end{pmatrix} = \begin{pmatrix} q_1(t) \\ q_3(t) \end{pmatrix}. \quad (55)$$

Together, the coarse-grained DG model becomes the LCM with material properties evaluated at the average temperature,

$$\begin{pmatrix} \overline{\rho c}_{p,1}|\Omega_1| & 0 \\ 0 & \overline{\rho c}_{p,2}|\Omega_2| \end{pmatrix} \begin{pmatrix} \dot{\bar{u}}_1 \\ \dot{\bar{u}}_2 \end{pmatrix} = \begin{pmatrix} -\frac{1}{R_{12}} & \frac{1}{R_{12}} \\ \frac{1}{R_{12}} & -\frac{1}{R_{12}} \end{pmatrix} \begin{pmatrix} \bar{u}_1 \\ \bar{u}_2 \end{pmatrix} + \begin{pmatrix} q_1(t) \\ q_3(t) \end{pmatrix} \quad (56)$$

## B. Appendix: Implementation Details

### A. Lumped Capacitance Model

Let $v_i$ represent the area of the $i$-th element, $\overline{\rho c}_{p,i}$ the heat capacity evaluated using the average temperature in the $i$-th element, and $1/R_{ij} = 1/R_i(\bar{u}_i) + 1/R_j(\bar{u}_j)$ the equivalent thermal resistance between elements $i$ and $j$. The LCM matrices for the LCM-3 and LCM-5 models are provided in Eqs. (57b) and (58b), respectively.

$$\bar{\mathbf{A}}(\bar{\mathbf{u}}) = \begin{pmatrix} \overline{\rho c}_{p,1} v_1 & 0 & 0 \\ 0 & \overline{\rho c}_{p,2} v_2 & 0 \\ 0 & 0 & \overline{\rho c}_{p,3} v_3 \end{pmatrix} \tag{57a}$$

$$\bar{\mathbf{B}}(\bar{\mathbf{u}}) = \begin{pmatrix} -\frac{1}{R_{12}} - \frac{1}{R_{13}} & \frac{1}{R_{12}} & \frac{1}{R_{13}} \\ \frac{1}{R_{12}} & -\frac{1}{R_{12}} - \frac{1}{R_{23}} & \frac{1}{R_{23}} \\ \frac{1}{R_{13}} & \frac{1}{R_{23}} & -\frac{1}{R_{13}} - \frac{1}{R_{23}} \end{pmatrix}, \quad \bar{\mathbf{f}}(t) = \begin{pmatrix} q_1(t) \\ q_2(t) \\ 0 \end{pmatrix} \tag{57b}$$

$$\bar{\mathbf{A}} = \begin{pmatrix} \overline{\rho c}_{p,1} v_1 & 0 & 0 & 0 & 0 \\ 0 & \overline{\rho c}_{p,2} v_2 & 0 & 0 & 0 \\ 0 & 0 & \overline{\rho c}_{p,3} v_3 & 0 & 0 \\ 0 & 0 & 0 & \overline{\rho c}_{p,4} v_4 & 0 \\ 0 & 0 & 0 & 0 & \overline{\rho c}_{p,5} v_5 \end{pmatrix} \tag{58a}$$

$$\bar{\mathbf{B}} = \begin{pmatrix} -\frac{1}{R_{12}} - \frac{1}{R_{15}} & \frac{1}{R_{12}} & 0 & 0 & \frac{1}{R_{15}} \\ \frac{1}{R_{12}} & -\frac{1}{R_{12}} - \frac{1}{R_{23}} - \frac{1}{R_{24}} & \frac{1}{R_{23}} & \frac{1}{R_{24}} & 0 \\ 0 & \frac{1}{R_{23}} & -\frac{1}{R_{23}} - \frac{1}{R_{34}} & \frac{1}{R_{34}} & 0 \\ 0 & \frac{1}{R_{24}} & \frac{1}{R_{34}} & -\frac{1}{R_{34}} - \frac{1}{R_{45}} & \frac{1}{R_{45}} \\ \frac{1}{R_{15}} & 0 & 0 & \frac{1}{R_{45}} & -\frac{1}{R_{45}} - \frac{1}{R_{15}} \end{pmatrix}, \quad \bar{\mathbf{f}}(t) = \begin{pmatrix} q_1(t) \\ q_2(t) \\ q_3(t) \\ 0 \\ 0 \end{pmatrix} \tag{58b}$$

## B. Operator-Inference Model

The Operator Inference (OpInf) framework is employed to construct a non-intrusive ROM for the transient thermal response of a multi-layered TPS. This approach learns low-dimensional operators governing the dynamical evolution of reduced modal coordinates obtained via proper orthogonal decomposition (POD), without requiring access to FOM equations. Let $\mathbf{\Pi} \in \mathbb{R}^{n \times r}$ be the POD basis computed from high-fidelity full-field temperature snapshots, where $n$ is the number of spatial degrees of freedom in the FOM, and $r \ll n$ is the reduced dimension. The reduced state is defined as $\mathbf{q}(t) = \mathbf{\Pi}^\top \mathbf{u}(t)$ where $\mathbf{u}(t)$ is the full-order temperature field. The structure of the model is informed by the high-fidelity discretization of the PDE as in Eq. (7),

$$\frac{d\mathbf{q}}{dt} = \mathbf{A}\mathbf{q}(t) + \mathbf{B}\mathbf{f}(t;\boldsymbol{\xi}_{BC}) + \mathbf{c}, \quad \mathbf{q}(t_0) = \boldsymbol{\Pi}^\top \mathbf{u}_0, \tag{59}$$

where $\mathbf{A} \in \mathbb{R}^{r \times r}$ is the reduced mass operator, $\mathbf{B} \in \mathbb{R}^{r \times m}$ captures the boundary condition effects via the parametrized input $\mathbf{f}(t;\boldsymbol{\xi}_{BC}) \in \mathbb{R}^m$, $\mathbf{c}$ is an affine offset term.

The OpInf is trained using the same dataset, $\mathcal{D}_1$, used for the PIROM listed in Sec. IV.B.2, except that the full-field measurement is used instead. The 20 full-order solutions in $\mathcal{D}_1$ are assembled over $K$ discrete time steps $\{t_k\}_{k=0}^{K-1}$ as reduced state $\mathbf{Q}_l$, input function $\mathbf{F}_l$, and time derivative $\mathbf{R}_l$ snapshots,

$$\mathbf{Q}_l = [\mathbf{q}(t_0), \ldots, \mathbf{q}(t_{K-1})] \in \mathbb{R}^{r \times K}, \quad \mathbf{F}_l = [\mathbf{f}(t_0), \ldots, \mathbf{f}(t_{K-1})] \in \mathbb{R}^{m \times K}, \quad \mathbf{R}_l = [\dot{\mathbf{q}}(t_0), \ldots, \dot{\mathbf{q}}(t_{K-1})] \in \mathbb{R}^{r \times K}$$

Thus, the known data is collected as $\mathbf{D}_l = \left[\mathbf{1}_k, \mathbf{Q}_l^\top, \mathbf{F}_l^\top\right] \in \mathbb{R}^{K \times d(r,m)}$ and the unknown operator is $\mathbf{O} = [\mathbf{c}, \mathbf{A}, \mathbf{B}] \in \mathbb{R}^{r \times d(r,m)}$, where $d(r,m) = (r+1)(r+2)/2 + m$, and $\mathbf{1}_k \in \mathbb{R}^K$ is a length-$K$ vector of 1's. Note that the problem parametrization appears only in $\mathbf{f}(t;\boldsymbol{\xi})$ for the boundary conditions in the $\mathcal{D}_1$ dataset.

The OpInf model is trained using the same computing resources used for the PIROM listed in Sec. IV.B.2. During training, a Tikhonov regularization strategy is introduced to mitigate overfitting. The regularization is necessary to produce a ROM that minimizes the objective function over the time domain of interest, while maintaining a bound on the integrated POD coefficients [11, 23]. Thus, the final learning problem for OpInf is,

$$\min_{\mathbf{O}} \sum_{l=1}^{N_s} \left\|\mathbf{D}_l \mathbf{O}^\top - \mathbf{R}_l^\top\right\|_2^2 + \left\|\boldsymbol{\Gamma}\mathbf{O}^\top\right\|_2^2, \tag{60}$$

where $\boldsymbol{\Gamma} = \text{diag}(\gamma_1 \mathbf{I}, \gamma_2 \mathbf{I}, \gamma_3 \mathbf{I})$ is the regularization matrix with penalties $\gamma_i > 0$ applied to each block of $\mathbf{O}^\top$. The OpInf learning problem results in an un-constrained least-squares minimization problem with $\mathbf{O}$ appearing linearly in the objective function.

The constants defining the size of the least-squares problem are defined as follows: (1) $K = 1,000$: empirically determined to provide a good balance between training computational cost and temporal resolution. (2) $r = 6$: number of POD modes used to represent the latent dynamics of the FOM, which yields 99.999% of the total energy of the system with a reconstruction error of $3.93 \times 10^{-7}$. (3) $m = 3$: total number of control inputs representing the boundary conditions on the three different layers. (4) $n = 5103$: number of spatial degrees of freedom in the FOM.

The optimal hyperparameter regularization is conducted using OpInf's model calibration algorithm `fit_regselect_continuous`. The ranges for the penalties $\gamma_1$, $\gamma_2$, and $\gamma_3$ are all log-spaced from $10^{-12}$ to $10^2$, with 10 points in each dimension. The model calibration algorithm reports the best regularization candidate via grid search and via optimization

as $\left[1 \times 10^2, 3.594 \times^{-11}, 7.743 \times 10^{-2}\right]$ and $\left[3.681 \times 10^3, 5.305 \times 10^{-12}, 1.763\right]$, respectively, with a fitting error of $\approx 0.25\%$. The total grid search and training computational costs amount to 54 minutes.

### C. Neural Ordinary Differential Equation

The Neural Ordinary Differential Equation (NODE) offers a purely data-driven model to approximate the dynamical evolution of temperature observables $\mathbf{z}(t) \in \mathbb{R}^3$ in the TPS. The model assumes the form,

$$\dot{\mathbf{z}} = \mathbf{g}_{NN}\left(\mathbf{z}, \mathbf{f}(t), t; \boldsymbol{\Theta}\right) \tag{61}$$

where $\mathbf{f} \in \mathbb{R}^3$ represents the time-varying boundary condition input, and $\boldsymbol{\Theta}$ are trainable neural network weights. The function $\mathbf{g}_{NN}$ is realized via a feedforward neural network with 7 hidden layers, 32-dimensional hidden states, and Parametric ReLU (PReLU) activation functions. Through a thorough cross-validation study, this architecture is empirically found to offer sufficient representational capacity to learn the TPS observable dynamics to the desired tolerances.

To train the NODE, the tolerance-sweeping algorithm (Alg. 1) is conducted on dataset $\mathcal{D}_1$, identical to that used for PIROM and OpInf. All hyperparameters are kept consistent with those used in PIROM training as detailed in Sec. IV.B.2, except for the learning rate, which is set to $10^{-3}$ to enable a faster, yet stable optimization. Optimization is performed using `ADAM`, and the adjoint-based gradients are computed using the `torchdiffeq` library.

## References

[1] Jenkins, D. R., *Space Shuttle: The History of the National Space Transportation System*, Voyageur Press, 2007.

[2] Price, A. B., "Design Report - Thermal Protection System X-15A-2," Tech. rep., National Aeronautics and Space Administration, 1968.

[3] Ohlhosrt, C. W., Glass, D., Bruce, W. E., lindell, M. C., Vaughn, W. L., Smith, R. W., Dirling, R. B., Hogenson, P. A., Nichols, J., Risner, N. W., Thompson, D. R., Kowbel, W., Sullivan, B. J., Koenig, J. R., and Cuneo, J. C., "Development of X-43A Mach 10 Leading Edges," *56th International Astronautical Congress of the International Astronautical Federation and The International Academy of Astronautics and The International Institute of Space Law*, Fukuoka Japan, 2005. https://doi.org/10.2514/6.IAC-05-D2.5.06.

[4] Beck, R. A. S., Driver, D. M., Wright, M. J., and Hwang, H. H., "Development of the Mars Science Laboratory Heatshield Thermal Protection System," *Journal of Spacecraft and Rockets*, Vol. 51, No. 4, 2014. https://doi.org/10.2514/1.A32635.

[5] Rasmussen, C. E., and Williams, C. K. I., *Gaussian Processes for Machine Learning*, The MIT Press, 2005. https://doi.org/10.7551/mitpress/3206.001.0001.

[6] Chen, R. T. Q., Rubanova, Y., Bettencourt, J., and Duvenaud, D., "Neural Ordinary Differential Equations," *arXiv*, 2018. https://doi.org/10.48550/arXiv.1806.07366.

[7] Sirovich, L., "Turbulence and the Dynamics of Coherent Structures Part I: Coherent Structures," *Quarterly of Applied Mathematics*, Vol. 45, No. 3, 1987, pp. 561–571.

[8] Sirovich, L., "Turbulence and the Dynamics of Coherent Structures Part II: Symmetries and Transformations," *Quarterly of Applied Mathematics*, Vol. 45, No. 3, 1987, pp. 573–582.

[9] Sirovich, L., "Turbulence and the Dynamics of Coherent Structures Part III: Dynamics and Scaling," *Quarterly of Applied Mathematics*, Vol. 45, No. 3, 1987, pp. 583–590.

[10] Willcox, K., and Peraire, J., "Balanced Model Reduction via the Proper Orthogonal Decomposition," *AIAA Journal*, Vol. 40, No. 11, 2012, pp. 2323–2330. https://doi.org/10.2514/2.1570.

[11] Benner, P., Goyal, P., Kramer, B., Peherstorfer, B., and Willcox, K., "Operator Inference for Non-Intrusive Model Reduction of Systems with Non-Polynomial Nonlinear Terms," *Computer Methods in Applied Mechanics and Engineering*, Vol. 372, 2020, p. 113433. https://doi.org/10.1016/j.cma.2020.113433.

[12] Arienti, M., Blonigan, P. J., Rizzi, F., Tencer, J., and Howard, M., "Projection-Based Model Reduction for Finite-Element Simulations of Thermal Protection Systems," *AIAA SciTech Forum*, Virtual, 2021. https://doi.org/10.2514/6.2021-1717.

[13] Blonigan, P. J., Tencer, J. T., and Rizzi, F., "Projection-based Reduced-Order Models with Hyperreduction for Finite Element Simulations of Thermal Protection Systems," *AIAA AVIATION 2023 Forum*, San Diego, California, 2023. https://doi.org/10.2514/6.2023-3913.

[14] Karniadakis, G. E., Kevrekidis, I. G., Lu, L., Perdikaris, P., and Yang, L., "Physics-Informed Machine Learning," *ResearchGate*, 2021, pp. 1–19. https://doi.org/10.1038/s42254-021-00314-5.

[15] Lui, H. F. S., and Wolf, W. R., "Convoutional Neural Networks for the Construction of Surrogate Models for Fluid Flows," 2021. https://doi.org/10.2514/6.2021-1675.

[16] Hao, Z., Liu, S., Zhang, Y., Ying, C., Feng, Y., Su, H., and Zhu, J., "Physics-Informed Machine Learning: A Survey on Problems, Methods and Applications," *arXiv*, 2022. https://doi.org/10.48550/arXiv.2211.08064.

[17] Penwarden, M., Zhe, S., Narayan, A., and Kirby, R. M., "A metalearning approach for Physics-Informed Neural Networks (PINNs): Application to parameterized PDEs," *Journal of Computational Physics*, Vol. 477, 2023, p. 111912. https://doi.org/10.1016/j.jcp.2023.111912.

[18] Yasar, H. A., and Sevinc, O. K., "Physics-Informed Neural Networks for Enhanced Critical Heat Flux Prediction in Hypersonic Flows," *AIAA Aviation Forum and ASCEND Conference*, Las Vegas, Nevada, 2024. https://doi.org/10.2514/6.2024-4203.

[19] Wang, R., and Yu, R., "Physics-Guided Deep Learning for Dynamical Systems: A Survey," *arXiv*, 2021. https://doi.org/10.48550/arXiv.2107.01272.

[20] Mao, Z., Lu, L., Marxen, O., Zaki, T. A., and Karniadakis, G. E., "DeepM&Mnet for hypersonics: Predicting the coupled flow and finite-rate chemistry behind a normal shock using neural-network approximation of operators," *Journal of Computational Physics*, Vol. 447, 2021, p. 110698. https://doi.org/10.1016/j.jcp.2021.110698, URL https://doi.org/10.1016/j.jcp.2021.110698.

[21] Peyvan, A., and Kumar, V., "Fusion DeepONet: A Data-Efficient Neural Operator for Geometry-Dependent Hypersonic Flows on Arbitrary Grids," *arXiv:2501.01934*, 2025.

[22] Hao, Y., Clark Di Leoni, P., Marxen, O., Meneveau, C., Karniadakis, G. E., and Zaki, T. A., "Instability-wave prediction in hypersonic boundary layers with physics-informed neural operators," *Journal of Computational Science*, Vol. 73, 2023, p. 102120. https://doi.org/10.1016/j.jocs.2023.102120.

[23] McQuarrie, S., C., S. H., and Willcox, K., "Data-Driven Reduced-Order Models Via Regularized Operator Inference for a Single-Injector Combustion Process," *Journal of the Royal Society of New Zealand*, 2021, pp. 194–211. https://doi.org/10.1080/03036758.2020.1863237.

[24] Blonigan, P. J., Tencer, J., Babiniec, S., and Murray, J., "Operator Inference-Based Model Order Reduction of Thermal Protection System Finite Element Simulations," *AIAA SCITECH 2025 Forum*, Orlando, Florida, 2025. https://doi.org/10.2514/6.2025-2133.

[25] Venegas, C. V., and Huang, D., "Expedient Hypersonic Aerothermal Prediction for Aerothermoelastic Analysis Via Field Inversion and Machine Learning," Virtual Event, 2021. https://doi.org/10.2514/6.2021-1707.

[26] Venegas, C. V., and Huang, D., "Physics-Infused Reduced Order Modeling of Hypersonic Aerothermal Loads for Aerothermoelastic Analysis," San Diego and Virtual, 2022. https://doi.org/10.2514/6.2022-0989.

[27] Venegas, C. V., and Huang, D., "Physics-Infused Reduced-Order Modeling of Aerothermal Loads for Hypersonic Aerothermoelastic Analysis," *AIAA Journal*, Vol. 61, No. 3, 2023. https://doi.org/10.2514/1.J062214.

[28] Venegas, C. V., and Huang, D., "Development of Weak-Form Physics-Infused Reduced Order Modeling With Applications," Orlando, FL, 2024. https://doi.org/10.2514/6.2024-0782.

[29] Yu, Y., Harlim, J., Huang, D., and Li, Y., "Learning Coarse-Grained Dynamics on Graph," *arXiv preprint arXiv:2405.09324*, 2024. https://doi.org/10.48550/arXiv.2405.09324.

[30] Parish, E., and Duraisamy, K., "A Unified Framework for Multiscale Modeling using the Mori-Zwanzig Formalism and the Variational Multiscale Method," 2017. https://doi.org/10.48550/arXiv.1712.09669.

[31] Parish, E., and Duraisamy, K., "Non-Markovian Closure Models for Large Eddy Simulations Using the Mori-Zwanzig Formalism," *Phys. Rev. Fluids*, Vol. 2, 2017, p. 014604. https://doi.org/10.1103/PhysRevFluids.2.014604.

[32] Stinis, P., "Higher-Order Mori-Zwanzig Models for the Euler Equations." *Multi-Scale Modeling and Simulation*, Vol. 6, No. 3, 2007, pp. 741–760. https://doi.org/10.48550/arXiv.math/0607108.

[33] Lin, Y. T., Tian, Y., Livescu, D., and Anghel, M., "Data-Driven Learning for the Mori–Zwanzig Formalism: A Generalization of the Koopman Learning Framework," *SIAM Journal on Applied Dynamical Systems*, Vol. 20, No. 4, 2021, pp. 2558–2601. https://doi.org/10.1137/21M1401759.

[34] Cohen, G., and Pernet, S., *Finite Element and Discontinuous Galerkin Methods for Transient Wave Equations*, Springer Dordrecht, 2018. https://doi.org/10.1007/978-94-017-7761-2.

[35] Incropera, F. P., DeWitt, D. P., Bergman, T. L., and Lavine, A. S., *Fundamentals of Heat and Mass Transfer*, 7th ed., Wiley, Hoboken, NJ, 2011.

[36] Parish, E., and Duraisamy, K., *Coarse-Graining Turbulence Using the Mori–Zwanzig Formalism*, Cambridge University Press, 2025.

[37] Betts, J. T., "Survey of Numerical Methods for Trajectory Optimization," *Journal of Guidance Control Dynamics*, Vol. 21, No. 2, 1998, pp. 193–207. https://doi.org/10.2514/2.4231.

[38] Klock, R. J., and Cesnik, C. E. S., "Nonlinear Thermal Reduced-Order Modeling for Hypersonic Vehicles," *AIAA Journal*, Vol. 55, No. 7, 2017, pp. 2358–2368. https://doi.org/10.2514/1.J055499.

[39] Clausen, J., Brunini, V., Collins, L., Knaus, R. C., Kucala, A., Lin, S., Moser, D. R., Phillips, M., Ransegnola, T. M., Subia, S. R., et al., "SIERRA Multimechanics Module: Aria Verification Manual-Version 5.22," Tech. rep., Sandia National Lab.(SNL-NM), Albuquerque, NM (United States), 2024.

[40] Paszke, A., Gross, S., Chintala, S., Chanan, G., Yang, E., DeVito, Z., Lin, Z., Desmaison, A., Antiga, L., and Lerer, A., "Automatic Differentiation in PyTorch," Tech. rep., October 2017.

[41] Kingma, D., and Ba, J., "Adam: A Method for Stochastic Optimization," *International Conference on Learning Representations*, 2014. https://doi.org/10.48550/arXiv.1412.6980.

[42] Yu, Y., Huang, D., Park, S., and Pangborn, H., "Learning Networked Dynamical System Models with Weak Form and Graph Neural Networks," *arXiv preprint arXiv:2407.16779*, 2024. https://doi.org/10.48550/arXiv.2407.16779.